\documentclass{emulateapj}
\slugcomment{to appear in ApJ. 2013 February 20, Issue 764-2:185} 
\usepackage{amsmath}

\usepackage{natbib}
\bibpunct{(}{)}{;}{a}{}{,}
\usepackage{graphicx}
\usepackage{amsmath}

\shorttitle{Distribution of High-Mass X-ray Binaries in the Milky Way}
\shortauthors{Coleiro \& Chaty}

\begin{document}

\title{Distribution of High-Mass X-ray Binaries in the Milky Way}

\author{Alexis Coleiro\altaffilmark{1} and Sylvain Chaty\altaffilmark{1,2}}

\altaffiltext{1}{Laboratoire AIM (UMR-E 9005 CEA/DSM - CNRS - Universit\'e Paris Diderot),
Irfu / Service d'Astrophysique, CEA-Saclay, 91191 Gif-sur-Yvette Cedex, France.}
\altaffiltext{2}{Institut Universitaire de France, 103 boulevard Saint Michel, 75005 Paris, France}

\email{alexis.coleiro@cea.fr; chaty@cea.fr}

\begin{abstract}
Observations of the high energy sky, particularly with the \textit{INTEGRAL} satellite, have quadrupled the number of supergiant X-ray Binaries observed in the Galaxy, and revealed new populations of previously hidden High Mass X-ray Binaries (HMXBs), raising new questions about their formation and evolution. The number of detected HMXBs of different types is now high enough to allow us to carry out a statistical analysis of their distribution in the Milky Way.
For the first time, we derive the distance and absorption of a sample of HMXBs using a Spectral Energy Distribution fitting procedure, and we examine the correlation with the distribution of Star Forming Complexes (SFCs) in the Galaxy. We show that HMXBs are clustered with SFCs with a typical cluster size of 0.3 $\pm$ 0.05 kpc and a characteristic distance between clusters of 1.7 $\pm$ 0.3 kpc. Furthermore, we present an investigation of the expected offset between the position of spiral arms and HMXBs, allowing us to constrain age and migration distance due to supernova kick for 13 sources. These new methods will allow us to assess the influence of the environment on these high energy objects with unprecedented reliability. 
\end{abstract}

\keywords{Galaxy: stellar content -- Stars: distances, early-type, evolution, formation -- X-rays: binaries}

\section{Introduction}

X-ray binaries are separated in two classes: Low Mass X-ray Binaries (LMXBs) and High Mass X-ray Binaries (HMXBs), differing by the mass of the companion star and the way accretion of matter occurs. HMXBs are binary systems composed of a compact object (neutron star or black hole candidate), accreting matter from a massive companion star, either a main sequence Be star or an evolved supergiant O or B star. Be stars are surrounded by a circumstellar decretion disc of gas, and accretion occurs periodically when the compact object passes through this disc of matter, whereas supergiant X-ray binaries are mostly wind-fed systems (see \citealt{Chaty_2011} for a review about these different kinds of HMXBs). Most of these sources are observed in the Galactic plane \citep{Grimm_2002} as expected since short-lived stellar systems do not have time to migrate far from their birthplace.\\
Due to dedicated observations with \textit{RXTE} and \textit{INTEGRAL}, around 200 HMXBs are currently known in the Milky Way allowing us to focus on their distribution. Using \textit{RXTE} data, \citet{Grimm_2002} highlighted clear signatures of the spiral structure in the spatial distribution of HMXBs. In the same way, \citet{Dean_2005}, \citet{Lutovinov_2005}, \citet{Bodaghee_2007} and \citet{Bodaghee_2012} showed that HMXBs observed with \textit{INTEGRAL} also seem to be associated with the spiral structure of the Galaxy. However, HMXB locations, mostly derived from their X-ray luminosity, are not well constrained and highly uncertain due to direct accretion occurring in HMXBs, well below the Eddington limit.\\
In order to overcome this caveat we present a comprehensive and novative approach allowing us to derive the locations of a sample of HMXBs, and study their distribution in the Galaxy.
We build the Spectral Energy Distribution (SED) of each HMXB and fit it with a blackbody model to compute the distance of each source.
We study this distribution and the correlation with Star Forming Complexes (SFCs) observed in the Galaxy. Knowing the location of these sources, one can examine the composition of the environment at their birthplace. This study, whose preliminary results have been presented in \citet{Coleiro_2011}, is necessary to better understand the formation and evolution of the whole population of HMXBs and primarily to state the role of the environment and binarity in the evolution of these high energy binary systems as outlined in the last section of this paper. In Section 2 we explain how we derive HMXB distances, then, in Section 3 we show that HMXBs are correlated with Star Forming Complexes. In Section 4, we derive the expected offset between HMXBs and Galactic spiral arms before discussing some implications on HMXB formation and evolution. We also derive the age and kick migration distance for 13 sources. Finally, we conclude in Section \ref{conclusion}.

\section{Deriving the HMXB location within our Galaxy}

To compute HMXB distances, we gathered a sample of HMXBs for which at least 4 optical and/or near infrared (NIR) magnitudes were known. For each source we build its SED and fit it with a blackbody model. This enabled us to evaluate the distance of the source along with its associated uncertainty.\\

There are currently more than 200 HMXBs detected in the Milky Way. Using the \citet{Liu_2006} catalogue, updated with literature, and the \textit{INTEGRAL} source catalogue of \citet{Bird_2010}, we retrieved the quiescent optical and NIR magnitudes (mostly from the 2MASS point source catalog), the spectral type and the luminosity class of each source from the literature. For each HMXB, four magnitudes are required for the fitting procedure, since the condition $N-n \ge 2$ with $N$ the number of observed magnitudes and $n$ the number of free parameters (in our study, $n=2$, cf. section \ref{fit}) needs to be met for $\chi^2$ statistics. Finally, 59 sources follow these requirements. 9 of them appear to be located in the Magellanic Clouds and four others are poorly known systems (IR counterparts are debated or physical parameters were not found in the literature) so our final sample consists of 46 sources in our Galaxy which meet the previous conditions (see Table \ref{sample1}).

\begin{center}
\begin{table*}
\scalebox{0.8}
{
\begin{tabular}{lllllll}

NAME & \textbf{D (kpc)} & \textbf{A$_{\rm{V}}$ (mag)} & T (K) & R (R$_{\sun}$) & SpT & SpT Ref.\\ \hline
\hline
1A 0535+262 & \textbf{3.8 $\pm$ 0.33} &\textbf{1.9 $\pm$ 0.26} & 32930 & 14.7  	& 	O9.7IIIe (O9.5IIIe) & \citet{Giangrande_1980}	 \\ 
1A 1118-615 & \textbf{3.2 $\pm$ 1.4} & \textbf{4.6 $\pm$ 1.9}      & 32930/34620 & 14.7/8.50 & 	O9.5IIIe/O9.5Ve (O9.5III/O9.5V)  & \citet{Janot-Pacheco_1981}	  \\ 
1E 1145.1-6141 & \textbf{10.5 $\pm$ 0.90}  & \textbf{5.2 $\pm$ 0.19} 	& 18300 & 51.0 &  B2Iae (B2Ia) & \citet{Densham_1982}	    \\ 
1H 1249-637 & \textbf{0.63 $\pm$ 2.5} & \textbf{1.8 $\pm$ 2.5} 	& 30160 & 14.8 &   B0.5IIIe (B0.5III)	 & \citet{Codina_1984}      \\ 
1H 1555-552 & \textbf{0.89 $\pm$ 0.093} &  \textbf{2.7 $\pm$ 0.6} 	&   19500 & 9.80 &      B2IIIn (B2III) & \citet{Liu_2006}	 	   \\ 
3A 0114+650 & \textbf{6.5 $\pm$ 3.0} & \textbf{4.0 $\pm$ 0.60}  & 24000 $\pm$ 3000 & 37.0 $\pm$ 15.0 & 	B1Ia   & \citet{Reig_1996}  		    \\ 
3A 0726-260 & \textbf{5.0 $\pm$ 0.82} & \textbf{2.5 $\pm$ 0.25}    &       38450/35900 & 9.30/8.80 & O8Ve/O9Ve (O8V/O9V) & \citet{Negueruela_1996}	      \\ 
3A 2206+543 & \textbf{3.4 $\pm$ 0.35} & \textbf{1.8 $\pm$ 0.60} &	  34620 & 8.50 &      	O9.5Vp (O9.5V) & \citet{Negueruela_2001_1} \\ 
4U 1700-377 & \textbf{1.8 $\pm$ 0.15} & \textbf{2.0 $\pm$ 0.15}  	   &  40210 &  21.2 & O6.5Iaf (O6.5Ia) & \citet{Wolff_1974}		\\ 
Cep X-4 & \textbf{3.7 $\pm$ 0.52} &   \textbf{5.3 $\pm$ 1.4} 	   & 22600/20500 & 6.17/5.62 &   B1Ve/B2Ve (B1V/B2V) & \citet{Bonnet-Bidaud_1998}	  \\ 
Cyg X-1 & \textbf{1.8 $\pm$ 0.56} & \textbf{3.4 $\pm$ 0.18}  & 32000 & 17.0 &   O9.7Iab & 	\citet{Walborn_1973}   \\ 
EXO 0331+530 & \textbf{6.9 $\pm$ 0.71} & \textbf{6.0 $\pm$ 0.50}   	    & 37170 & 9.00 &   O8.5Ve (O8.5V)  & 	\citet{Negueruela_1998} 	    \\ 
EXO 2030+375 & \textbf{3.1 $\pm$ 0.38} &  \textbf{12 $\pm$ 1.4}  	    & 33340 & 8.30 &   B0Ve (B0V)	  & 	\citet{Liu_2006}     \\ 
gam Cas & \textbf{0.17 $\pm$ 0.50} &   \textbf{1.2 $\pm$ 2.2} &	 25000-30000 & 10.0 &      	B0.5IVe		& \citet{Stee_1995}			   \\ 
GRO J1008-57 & \textbf{4.1 $\pm$ 0.59} &  \textbf{6.7 $\pm$ 1.1}  	    & 33340 & 8.30 &   B0e (B0V) & \citet{Belczynski_2009}	   	      	        \\ 
GT 0236+610 & \textbf{1.8 $\pm$ 0.20} &\textbf{3.8 $\pm$ 0.65}  	 & 333340 & 8.30 &     B0Ve (B0V)	   &   \citet{Crampton_1978_1}		 \\   
GX 301-2 & \textbf{3.1 $\pm$ 0.64} & \textbf{6.3 $\pm$ 0.14}  & 20400 & 62.0 &   B1Ia & \citet{Hammerschlag-Hensberge_1979}	      \\ 
GX 304-1 & \textbf{1.3 $\pm$ 0.10} & \textbf{6.0 $\pm$ 0.17} & 	20500 & 5.62 & 	     B2Vne  (B2V)	& \citet{Parkes_1980}	      \\ 
Ginga 0834-430 & \textbf{7.1 $\pm$ 4.2} &   \textbf{11 $\pm$ 2.2 } 	  & 31540/33340/ & 14.7/8.30/ &	B0/2IIIe/Ve (B0/2III/V)	 & \citet{Israel_2000}    	        \\ 
 & & &19500/20500 & 9.77/5.62 & &\\
H 0115+634 & \textbf{5.3 $\pm$ 0.44} & \textbf{6.4 $\pm$ 0.28}  & 	 33340 & 8.30 & B0.2Ve (B0V)	& \citet{Negueruela_2001}	           \\ 
H 1145-619 & \textbf{4.3 $\pm$ 0.52} &  \textbf{1.7 $\pm$ 0.55}  	   	& 31540 & 14.7 &  B0.2IIIe (B0III)	& \citet{Okazaki_2001}	      \\ 
H 1417-624 & \textbf{7.0 $\pm$ 0.74} & \textbf{6.1 $\pm$ 1.1} 	& 22600 & 6.17 &  B1Ve (B1V)	& \citet{Belczynski_2009}		       \\ 
H 1538-522 & \textbf{6.2 $\pm$ 1.8} & \textbf{6.4 $\pm$ 0.28}   	    & 28000 $\pm$ 2000 & 17.0 $\pm$ 2.00 & 	B0Iab	& \citet{Crampton_1978}	        \\ 
&&&31500 $\pm$ 1000 & 17.0 $\pm$ 2.00 &&\\
IGR J00370+6122 & \textbf{3.4 $\pm$ 1.2} &  \textbf{2.4 $\pm$ 0.19}  	& 24300/30160 & 21.75/14.8 &  	B0.5II/III	& \citet{Negueruela_2004}		\\ 
IGR J01583+6713 & \textbf{4.1 $\pm$ 0.63} & \textbf{4.7 $\pm$ 0.32} & 20000 & 7.70 & B2IVe & \citet{Kaur_2008}		  \\ 
IGR J06074+2205 & \textbf{4.5 $\pm$ 0.36} & \textbf{3.3 $\pm$ 0.22}  	  	& 32060 & 8.00 & 	B0.5Ve (B0.5V) & \citet{Reig_2010}		    \\ 
IGR J08408-4503 & \textbf{3.4 $\pm$ 0.35} & \textbf{1.7 $\pm$ 0.60}  	   	& 34230 & 23.8 & 	O8.5Ib(f) (O8.5I)	& \citet{Barba_2006}	     \\ 
IGR J11215-5952 & \textbf{7.3 $\pm$ 0.68}&  \textbf{2.6 $\pm$ 0.60}  	    	& 22000 & 36.5 &	B1Ia		& \citet{Liu_2006}      \\ 
IGR J11305-6256 & \textbf{3.6 $\pm$ 0.71}&  \textbf{1.0 $\pm$ 2.2}    & 31540 & 14.7 & B0IIIe (B0III) & \citet{Tomsick_2008}					       \\ 
IGR J11435-6109 & \textbf{9.8 $\pm$ 0.86} &  \textbf{5.7 $\pm$ 0.28} & 	 32060 & 8.00 &      	B0.5Ve (B0.5V)	& \citet{Torrejon_2004}			       \\ 
IGR J16465-4507 & \textbf{12.7 $\pm$ 1.3} & \textbf{5.0 $\pm$ 0.75} &	    	23600 & 33.1 & B0.5I		& \citet{Rahoui_2008}	        \\ 
IGR J17200-3116 & \textbf{10.4 $\pm$ 3.6} & \textbf{6.6 $\pm$ 4.5} 	    &  32060 & 8.00 & 	B0.5Ve (B0.5V)	& Coleiro et al. (in preparation)\\ 
IGR J18214-1318 & \textbf{10.6 $\pm$ 5.0} & \textbf{14 $\pm$ 3.1} 	   	& 32740 & 24.6 &  O9I & \citet{Butler_2009}				   \\ 
IGR J18410-0535 & \textbf{7.8 $\pm$ 0.74} & \textbf{6.1 $\pm$ 0.65} 	   	& 21700 & 34.9 &  B1Ib & \citet{Nespoli_2007}				    \\ 
IGR J18450-0435 & \textbf{6.4 $\pm$ 0.76} & \textbf{6.7 $\pm$ 0.49} & 	  31240 & 25.4 &  O9.5I		& \citet{Zurita-Heras_2009}	    \\ 
KS 1947+300 & \textbf{8.5 $\pm$ 2.3} & \textbf{4.0 $\pm$ 0.49} 		& 33340 & 8.30 & B0Ve (B0V)	& \citet{Negueruela_2003}		     \\ 
PSR B1259-63 & \textbf{1.7 $\pm$ 0.56} & \textbf{3.8 $\pm$ 0.70}	  & 32000$^{+2000}_{-1000}$ & 9.00$^{+1.8}_{-1.5}$ &	B2Ve	& \citet{Johnston_1994}		      \\   
RX J0440.9+4431 & \textbf{2.9 $\pm$ 0.37} & \textbf{2.9 $\pm$ 0.25}    	& 33340 & 8.30 & 	B0.2Ve (B0V)	 & \citet{Reig_2005}				  \\ 
RX J0812.4-3114 & \textbf{8.6 $\pm$ 1.8} & \textbf{2.3 $\pm$ 0.20}  	  	& 28000 $\pm$ 2000 & 10 $\pm$ 2.0 &  B0.2IVe		&  \citet{Reig_2001}			   \\ 
RX J1744.7-2713 & \textbf{1.2 $\pm$ 0.46} & \textbf{2.7 $\pm$ 0.57}  	  	& 30160/32060 & 14.8/8.00 &  B0.5IIIe/Ve (B0.5III/V)		& 	\citet{Liu_2006}		    \\ 
SAX J1818.6-1703 & \textbf{2.7 $\pm$ 0.28} &  \textbf{1.6 $\pm$  0.80}   & 17000 & 8.71 & 	B3III		& 	\citet{Liu_2006}			    \\ 
SAX J2103.5+4545 & \textbf{8.0 $\pm$ 0.78} & \textbf{4.2 $\pm$ 0.25} & 	       33340 & 8.30 & 	B0Ve (B0V)		& \citet{Reig_2004}			     \\ 
Vela X-1 & \textbf{2.2 $\pm$ 0.22} & \textbf{2.2 $\pm$ 0.46}  & 24700 & 33.8 &  B0.5Iae (B0.5Ia)	& 	\citet{Prinja_2010}				      \\ 
XTE J1855-026 & \textbf{10.8 $\pm$ 1.0} &  \textbf{5.8 $\pm$ 0.90} 		& 28100 & 26.9 & B0Iaep (B0Ia)		& \citet{Negueruela_2008}			        \\ 
XTE J1946+274 & \textbf{6.2 $\pm$ 3.0} & \textbf{6.9 $\pm$ 0.74}   	 & 32430/22050/ & 11.5/8.10/ &  B0/1/IVe/Ve (B0/1/IV/V)		& 		\citet{Belczynski_2009}				 \\ 
&&&33340/22600&8.30/6.17&&\\
X Per & \textbf{1.2 $\pm$ 0.16} & \textbf{0.81 $\pm$ 0.22 } &	  	      33340 & 8.30 & 	B0Ve (B0V)	&  	\citet{Belczynski_2009}			  \\

\hline
\end{tabular}
}
 \caption{The HMXB sample. Distance D, Extinction A$_{\rm{V}}$, T, R, spectral type SpT and spectral type reference SpT Ref. are given. D and A$_ {\rm{V}} $ are computed in this article whereas T and R are derived from \citet{Vacca_1996}, \citet{Panagia_1973} and \citet{Searle_2008} or taken in the literature when available (see Table \ref{refRT}). When different spectral types are proposed in the literature, the different R and T values used are given. Spectral types between brackets are the spectral types used for R and T determination. D and A$_{\rm{V}}$ error determination are developed in Section \ref{unc_section}.}
 \label{sample1}

\end{table*}

\end{center}

\subsection{SED fitting procedure}\label{fit}

\begin{figure}[h]
\resizebox{\hsize}{!}{\includegraphics{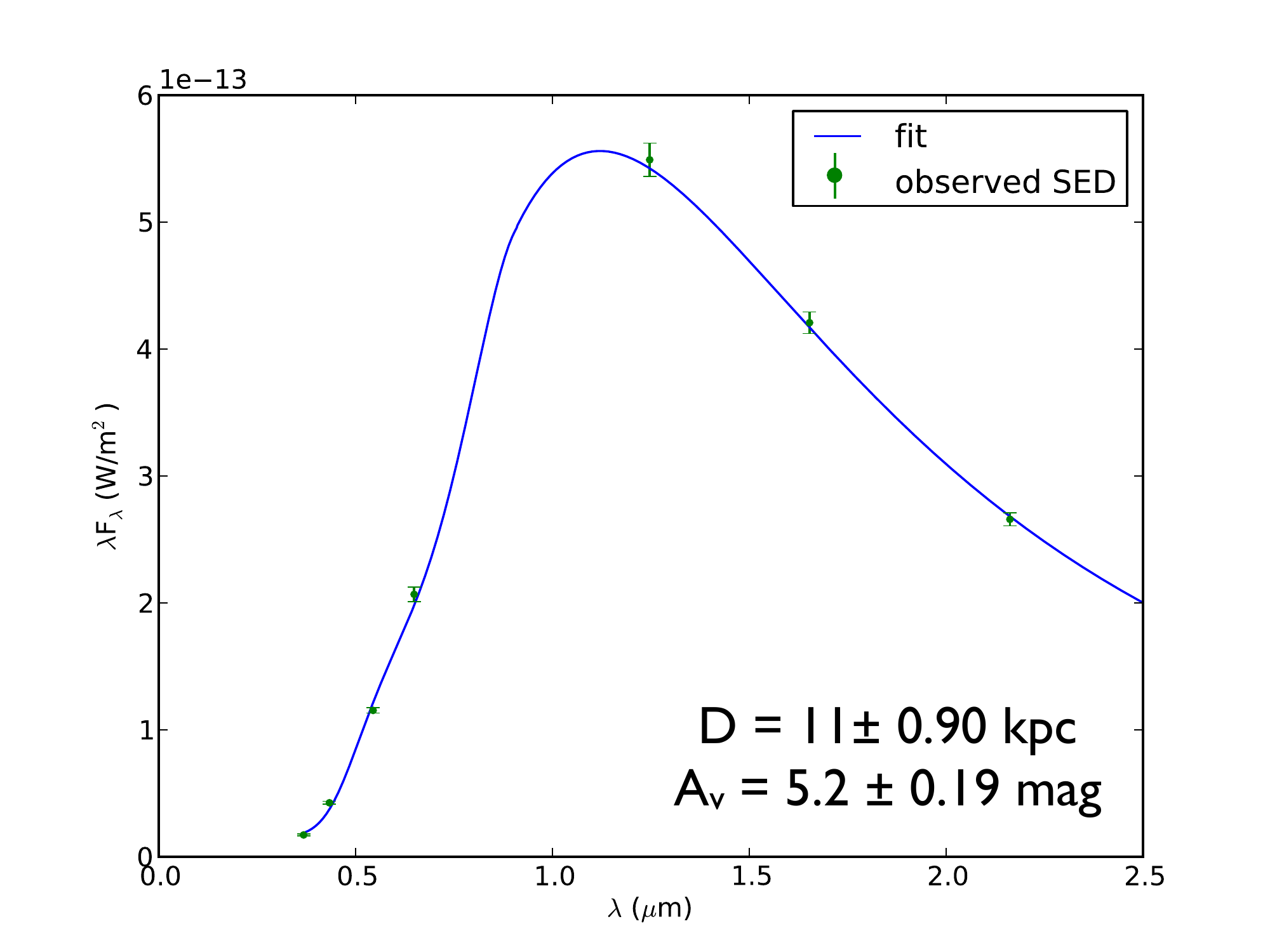}}
\caption{Result of the fitting for the source 1E 1145.1-6141 with distance D and extinction in V band $\rm{A}_{\rm{V}}$.}
\label{fits}
\end{figure}

Our fitting procedure is based on the Levenberg-Marquardt least-square algorithm implemented in Python (\verb leastsq  routine from the \verb scipy  package). For each HMXB, we build the SED in optical and NIR from a maximum of 8 (U, B, V, R, I, J, H, Ks) and a minimum of 4 magnitudes points. This SED is then fit (see Figure \ref{fits} for an example) by a blackbody model, with the aim of using an uniform fitting method for the entire sample of HMXBs, given by the relation:

\begin{equation}
\lambda F_{\lambda} =\frac{2\pi hc^2}{\lambda^4}\times10^{-0.4\rm{A}_{\lambda}}\frac{\left(\rm{R/D}\right)^2}{\exp{\left(\frac{hc}{\lambda k_{B}\rm{T}}\right)}-1} 
\label{blackbody}
\end{equation}

where: $\lambda$ is the wavelength in $\mu$m; $F_{\lambda}$ the flux density in W\,m$^{-2}$\,$\mu$m$^{-1}$; $h$ the Planck constant; $c$ the speed of light; $\rm{A}_\lambda$ the extinction at the wavelength $\lambda$; R/D the stellar radius over distance ratio and $k_B$ the Boltzmann constant and T the temperature of the star. When the radius and the temperature of the companion star are available in the literature, we use these values (given in Table \ref{refRT}). For the other sources, we derive the radius and the temperature of the companion star, which dominates the optical and NIR flux, from the spectral type and the luminosity class, by using tables of \citealt{Vacca_1996, Panagia_1973, Martins_2006, Searle_2008}, thereafter noted PVMS. Since only four sources (including GX 301-2) with a highly evolved mass star have been detected in the Milky Way \citep{Mason_2012}, we can reasonably assume that for all the other systems studied here, the radius of the companion star is sensibly close to the one expected, depending on the luminosity class. When the spectral type is poorly known, we derive the mean distance value by considering the different possible spectral types and we allocate adequate error on the distance of these systems. Finally, if no data is given in \citealt{Vacca_1996, Panagia_1973, Searle_2008} about a certain spectral type (which is the case for luminosity classes II and IV) we computed intermediate R and T between luminosity classes I and III for class II and between class III and V for class IV respectively. This affects only 3 sources in our sample.  For stars with peculiarities (\ion{N}{3} and \ion{He}{2} in emission mentioned by an 'f' in the spectral type, broad absorption mentioned by an 'n', unspecified peculiarity mentioned by a 'p') we assume their radius and temperature to have the same values than "normal" stars. For Be stars, we take R and T values of normal B stars, and the circumstellar disc emission is taken into account through a possible uncertainty due to the expected infrared excess of this family of sources (see Section \ref{unc_section}). Finally, we describe our estimate of distance uncertainties due to errors on R and T of companion star in Section \ref{unc_section}.

\begin{center}
\begin{table*}
\begin{center}

\scalebox{1}
{
\begin{tabular}{llll}

SOURCE NAME & R (R$_{\sun}$)  & T (K) & Reference\\
\hline
3A 0114+650 & 37.0 $\pm$ 15.0 & 24000 $\pm$ 3000 & \citet{Reig_1996}\\
Cyg X-1 & 17.0 & 32000  & \citet{Herrero_1995}\\
Gam Cas & 10.0 &  25000-30000 & \citet{Stee_1995} and \citet{Goraya_2007} \\
GX 301-2 & 62.0 & 20400 & \citet{Kaper_2006}\\
H 1538-522 (ref. 1)& 17.2 $\pm$ 1 & 28000 $\pm$ 2000 & \citet{Reynolds_1992}\\
H 1538-522 (ref. 2)&17 $\pm$ 2.0 & 31500 $\pm$ 1000 & \citet{Crampton_1978} \\
PSR B1259-63 & 9.0$^{+1.8}_{-1.5}$ & 32000$^{+2000}_{-1000}$ & \citet{Negueruela_2011}\\
RX J0812.4-3114 & 10 $\pm$ 2 & 28000 $\pm$ 2000 & \citet{Reig_2001}	\\
\hline

\end{tabular}
}
\caption{Radius R and Temperature T values available in the literature with references.}
\label{refRT} 
\end{center}
\end{table*}
\end{center}


Moreover, two parameters are left free: the extinction in V band $\rm{A}_{\rm{V}}$ and the ratio R/D, whereas the extinction $A_\lambda$ is derived at each wavelength from \citet{Cardelli_1989} assuming $R_{\rm{V}}=3.1$ (corresponding to the mean value derived for a diffuse interstellar medium towards the Galactic plane, \citealt{Fitzpatrick_1999}). The influence of potential change of the extinction law was explored. The results, presented in Section \ref{unc_section}, do not show any substantial variation in distance determination due to extinction change.\\

Knowing the radius R of the companion star, we then calculate the distance D in kpc.\\

The least square function given by the formula $\chi^2=\sum_i\left[\frac{X_{i,obs}-X_{i,model}}{\sigma_i}\right]^2$ (with $X_{i,obs}$ the observed flux value for the filter $i$, $X_{i,model}$, the predicted flux in the i$^{th}$ filter derived from the black-body model and $\sigma_i$, the flux error in the same filter) is then minimized by the Levenberg-Marquardt algorithm.\\

To check our results, we estimated the distance of each HMXB using the expression

\begin{equation}
\rm{D}_{\rm{pc}} = 10^{-0.4(\rm{m}_{\rm{V}}-\rm{M}_{\rm{V}}-\rm{A}_{\rm{V}}+5)}
\label{distance2}
\end{equation}

with $\rm{m}_{\rm{V}}$ the relative magnitude and $\rm{M}_{\rm{V}}$ the absolute magnitude while $\rm{A}_{\rm{V}}$ is given by the fit. This formula has the advantage of not depending on the companion star radius. However, it does depend on the absolute magnitude M$_{\rm{V}}$, given in \citet{Martins_2006} for O type stars and in \citet{Morton_1968} and \citet{Panagia_1973} for B stars. The results obtained with this second method are consistent with those using the first approach.\\

We give results in Table \ref{sample1} and compare these results with previously published results in Appendix \ref{annex_comp}. We point out that median discrepancy in distance is $\sim 17 \%$ and $\rm{A}_{\rm{V}}$ is often very similar (median discrepancy of less than 7\%).

\begin{center}
\begin{table*}
\scalebox{1}
{
\begin{tabular}{llllll}

NAME & D (kpc) & A$_{\rm{V}}$ (mag) & D$_{\rm{ref}}$ (kpc) & A$_{\rm{Vref}}$ (mag) & D and A$_{\rm{V}}$ Ref.\\ \hline
\hline
1A 0535+262 & \textbf{3.8 $\pm$ 0.33} &\textbf{1.9 $\pm$ 0.26} & 1.8 $\pm$ 0.6 & 2.7 $\pm$ 0.2  	 & \citet{Giangrande_1980}	 \\ 
1A 1118-615 & \textbf{3.2 $\pm$ 1.4} & \textbf{4.6 $\pm$ 1.9}      & 3.6 $\pm$ 0.9 & 4.2 max  & \citet{Janot-Pacheco_1981}	  \\ 
1E 1145.1-6141 & \textbf{10.5 $\pm$ 0.90}  & \textbf{5.2 $\pm$ 0.19} 	& 8.2 $\pm$ 1.5 & -- & \citet{Densham_1982}	    \\ 
1H 1249-637 & \textbf{0.63 $\pm$ 2.5} & \textbf{1.8 $\pm$ 2.5} 	& 0.43 $\pm$ 0.060 & --  & \citet{Codina_1984}      \\ 
& & & 0.30$^{max=0.36}_{min=0.26}$ & --  & \citet{Chevalier_1998}\\
3A 0114+650 & \textbf{6.5 $\pm$ 3.0} & \textbf{4.0 $\pm$ 0.60}  &  7.0 $\pm$ 3.6 & 	 --  & \citet{Reig_1996}  		    \\ 
3A 0726-260 & \textbf{5.0 $\pm$ 0.82} & \textbf{2.5 $\pm$ 0.25}    &       6.1 $\pm$ 0.30 & -- &  \citet{Negueruela_1996}	      \\ 
3A 2206+543 & \textbf{3.4 $\pm$ 0.35} & \textbf{1.8 $\pm$ 0.60} &	  2.6 & -- & \citet{Blay_2006} \\ 
4U 1700-377 & \textbf{1.8 $\pm$ 0.15} & \textbf{2.0 $\pm$ 0.15}  	   &  1.7 &  --  & \citet{Ankay_2001}		\\ 
Cep X-4 & \textbf{3.7 $\pm$ 0.52} &   \textbf{5.3 $\pm$ 1.4} 	   & 3.8 $\pm$ 0.60 & -- & \citet{Bonnet-Bidaud_1998}	  \\ 
Cyg X-1 & \textbf{1.8 $\pm$ 0.56} & \textbf{3.4 $\pm$ 0.18}  & 1.86 $^{+0.12}_{-0.11}$ &  -- & 	\citet{Reid_2011}   \\ 
EXO 0331+530 & \textbf{6.9 $\pm$ 0.71} & \textbf{6.0 $\pm$ 0.50}   	    &  6 $<$ d $<$ 9 &   --  & 	\citet{Negueruela_1999} 	    \\ 
EXO 2030+375 & \textbf{3.1 $\pm$ 0.38} &  \textbf{12 $\pm$ 1.4}  	    & 7.1 $\pm$ 0.20 & & 	\citet{Wilson_2002}     \\ 
gam Cas & \textbf{0.17 $\pm$ 0.50} &   \textbf{1.2 $\pm$ 2.2} &	 0.188$^{max=0.208}_{min=0.168}$ &      	--		& \citet{Chevalier_1998}			   \\ 
GRO J1008-57 & \textbf{4.1 $\pm$ 0.59} &  \textbf{6.7 $\pm$ 1.1}  	    & 5 & -- & \citet{Coe_1994}	   	      	        \\ 
GT 0236+610 & \textbf{1.8 $\pm$ 0.20} &\textbf{3.8 $\pm$ 0.65}  	 & 2 & --	   &   \citet{Steele_1998}		 \\   
GX 301-2 & \textbf{3.1 $\pm$ 0.64} & \textbf{6.3 $\pm$ 0.14}  & 3 -- 4  &   -- &  \citet{Kaper_2006}     \\ 
GX 304-1 & \textbf{1.3 $\pm$ 0.10} & \textbf{6.0 $\pm$ 0.17} & 	2.4 $\pm$ 0.50 & -- & \citet{Parkes_1980}	      \\ 
Ginga 0834-430 & \textbf{7.1 $\pm$ 4.2} &   \textbf{11 $\pm$ 2.2 } 	  & 3 $<$ d $<$ 5 &-- & \citet{Israel_2000}    	        \\ 
H 0115+634 & \textbf{5.3 $\pm$ 0.44} & \textbf{6.4 $\pm$ 0.28}  & 	7-8 & --	& \citet{Negueruela_2001}	           \\ 
H 1145-619 & \textbf{4.3 $\pm$ 0.52} &  \textbf{1.7 $\pm$ 0.55}  	   	& 3.1 &  --	& \citet{Stevens_1997}	      \\ 
H 1417-624 & \textbf{7.0 $\pm$ 0.74} & \textbf{6.1 $\pm$ 1.1} 	& 1.4 $<$ d $<$ 11.1 &  6.1 $<$ A$_{\rm{V}}$ $<$ 8.9 & \citet{Grindlay_1984}		       \\ 
H 1538-522 & \textbf{6.2 $\pm$ 1.8} & \textbf{6.4 $\pm$ 0.28}   	    & 5.5 $\pm$ 1.5 &  -- & \citet{Crampton_1978}	        \\ 
 &  &     & 6.4 $\pm$ 1.0 &  -- & \citet{Reynolds_1992}	        \\ 
 &  &     & 4.5  &  6.5 $\pm$ 0.3 & \citet{Clark_2004}	        \\ 
IGR J00370+6122 & \textbf{3.4 $\pm$ 1.2} &  \textbf{2.4 $\pm$ 0.19}  	& 3.0 &  	-- & \citet{Reig_2005}		\\ 
IGR J01583+6713 & \textbf{4.1 $\pm$ 0.63} & \textbf{4.7 $\pm$ 0.32} & 4.0 $\pm$ 0.4 & 4.5 $\pm$ 0.2 & \citet{Kaur_2008}		  \\ 
IGR J08408-4503 & \textbf{3.4 $\pm$ 0.35} & \textbf{1.7 $\pm$ 0.60}  	   	& 2.7 & --	& \citet{Leyder_2007}	     \\ 
IGR J11215-5952 & \textbf{7.3 $\pm$ 0.68}&  \textbf{2.6 $\pm$ 0.60}  	    	&8.0  &	--		& \citet{Negueruela_2005}      \\ 
IGR J11305-6256 & \textbf{3.6 $\pm$ 0.71}&  \textbf{1.0 $\pm$ 2.2}    & 3 & -- & \citet{Masetti_2006}\\					       
IGR J11435-6109 & \textbf{9.8 $\pm$ 0.86} &  \textbf{5.7 $\pm$ 0.28}  & 4.5 &     -- & \citet{Torrejon_2004}	 \\ 
IGR J16465-4507 & \textbf{12.7 $\pm$ 1.3} & \textbf{5.0 $\pm$ 0.75} &	    12.5  & -- & \citet{Smith_2004}	        \\ 
IGR J18214-1318 & \textbf{10.6 $\pm$ 5.0} & \textbf{14 $\pm$ 3.1} 	   	& 10 & -- & \citet{Butler_2009}				   \\ 
IGR J18410-0535 & \textbf{7.8 $\pm$ 0.74} & \textbf{6.1 $\pm$ 0.65} 	   	& 3.2 &  -- & \citet{Nespoli_2008}				    \\ 
IGR J18450-0435 & \textbf{6.4 $\pm$ 0.76} & \textbf{6.7 $\pm$ 0.49} & 3.6 &  --	& \citet{Zurita-Heras_2009}	    \\ 
KS 1947+300 & \textbf{8.5 $\pm$ 2.3} & \textbf{4.0 $\pm$ 0.49} 		& 10 & 3.38	& \citet{Negueruela_2003}		     \\ 
PSR B1259-63 & \textbf{1.7 $\pm$ 0.56} & \textbf{3.8 $\pm$ 0.70}	  & 2.3 &	--	& \citet{Negueruela_2011}		      \\   
RX J0440.9+4431 & \textbf{2.9 $\pm$ 0.37} & \textbf{2.9 $\pm$ 0.25}    	& 3.3 & 	--	 & \citet{Reig_2005}				  \\ 
RX J0812.4-3114 & \textbf{8.6 $\pm$ 1.8} & \textbf{2.3 $\pm$ 0.20}  	  	&8.8 &  --		&  \citet{Reig_2001}			   \\ 
RX J1744.7-2713 & \textbf{1.2 $\pm$ 0.46} & \textbf{2.7 $\pm$ 0.57}  	  	&0.8 $\pm$ 0.1 & --	& 	\citet{Motch_1997}		    \\ 
SAX J1818.6-1703 & \textbf{2.7 $\pm$ 0.28} &  \textbf{1.6 $\pm$  0.80}   & 2.5 & 	--	& 	\citet{Sidoli_2009}			    \\ 
SAX J2103.5+4545 & \textbf{8.0 $\pm$ 0.78} & \textbf{4.2 $\pm$ 0.25} & 	      6.5 & 4.2 $\pm$ 0.3	& \citet{Reig_2004}			     \\ 
Vela X-1 & \textbf{2.2 $\pm$ 0.22} & \textbf{2.2 $\pm$ 0.46}  &1.9 $\pm$ 0.1 & --	& 	\citet{Sadakane_1985}				      \\ 
XTE J1855-026 & \textbf{10.8 $\pm$ 1.0} &  \textbf{5.8 $\pm$ 0.90} 		&10 & --	& \citet{Corbet_1999}			        \\ 
XTE J1946+274 & \textbf{6.2 $\pm$ 3.0} & \textbf{6.9 $\pm$ 0.74}   	 &9.5 $\pm$ 2.9 &  --	& 		\citet{Wilson_2003}				 \\ 
X Per & \textbf{1.2 $\pm$ 0.16} & \textbf{0.81 $\pm$ 0.22 } &	 0.70 $\pm$ 0.30 & 1.05 $\pm$ 0.02  	&  	\citet{Lyubimkov_1997}			  \\

\hline
\end{tabular}
}
 \caption{Comparison of D and A$_{\rm{V}}$ values derived in this study with D$_{\rm{ref}}$ and A$_{\rm{Vref}}$ available in the literature.}
 \label{comp_table}

\end{table*}

\end{center}

\subsection{Uncertainties in the computed distance and extinction}\label{unc_section}
The magnitude uncertainties are retrieved from the literature. For sources for which no error is given, we use a systematic error of 0.1 magnitude, conservative of typical errors. The flux uncertainties are then derived from these magnitude errors. Nonetheless, we assume in the following that the spectral type given in the literature is the real spectral type of the companion star. We carried out simulations on a wide range of spectral types which enabled us to constrain the uncertainty that appears to be less important for supergiant stars than for Zero-Age Main Sequence stars.\\

Uncertainties on the radius and temperature of the companion star could have a severe influence on distance and $\rm{A}_{\rm{V}}$ error. We generate SEDs of various sources using a large range of temperatures which clearly show that the distance error is less sensible to the temperature error than to the radius one. That can be easily understood given the fact that distance is directly derived using the R/D ratio.  We distinguished different cases. 1) For sources for which spectral type is accurately determined, we select R and T values mostly in \citealt{Vacca_1996} for consistency. If no data is given in this reference, we used R and T values given in \citet{Panagia_1973}, \citet{Martins_2006} or \citet{Searle_2008}. With the aim of deriving accurate distance and $\rm{A}_{\rm{V}}$ errors due to radius and temperature uncertainties, we fitted the SED of each system using the R and T values given in PVMS and we compute the dispersion in distance and $\rm{A}_{\rm{V}}$ values obtained using these four references. These computations lead to a mean difference in distance value of 10\% as well as for the $\rm{A}_{\rm{V}}$ value. Then, for systems for which spectral type is well known, we consider a systematic error of 10\% of the derived distance and $\rm{A}_{\rm{V}}$ values due to radius and temperature uncertainties. 2) For sources for which values of R and T are available in the literature (see Table \ref{refRT}), we use these values in our fitting procedure to derive distance and extinction in V band. When maximum and minimum radius and temperature values are available, we derive distances and dispersion on distance using these values. We also derive the distance considering R and T values given in PVMS. Then, we consider as the final error (due to radius and temperature uncertainties) the dispersion on all the distance values obtained with R and T values given in the four references and with R and T values given in the paper dedicated to the source (referenced in Table \ref{refRT}). The same approach was used to derive an $\rm{A}_{\rm{V}}$ error due to radius and temperature uncertainties. 3) When the spectral type is not well constrained and if no R and T value is available in the literature, we derived distances (and then a mean distance) using the different possible spectral types of the companion star and then the different R and T values given in PVMS. We finally compute the dispersion on all the distances derived and considered it as the error due to radius and temperature uncertainties. We derive an $\rm{A}_{\rm{V}}$ error due to radius and temperature uncertainties for these sources, using the same method. 4) For luminosity classes II and IV, since no data are given in PVMS we used intermediate R and T values between class I and III for class II and  between class III and V for luminosity class IV. Errors were computed using the dispersion on distance obtained with these R and T values and distances obtained with R and T values of lower and higher luminosity classes.\\

\begin{figure}[h]
\resizebox{\hsize}{!}{\includegraphics{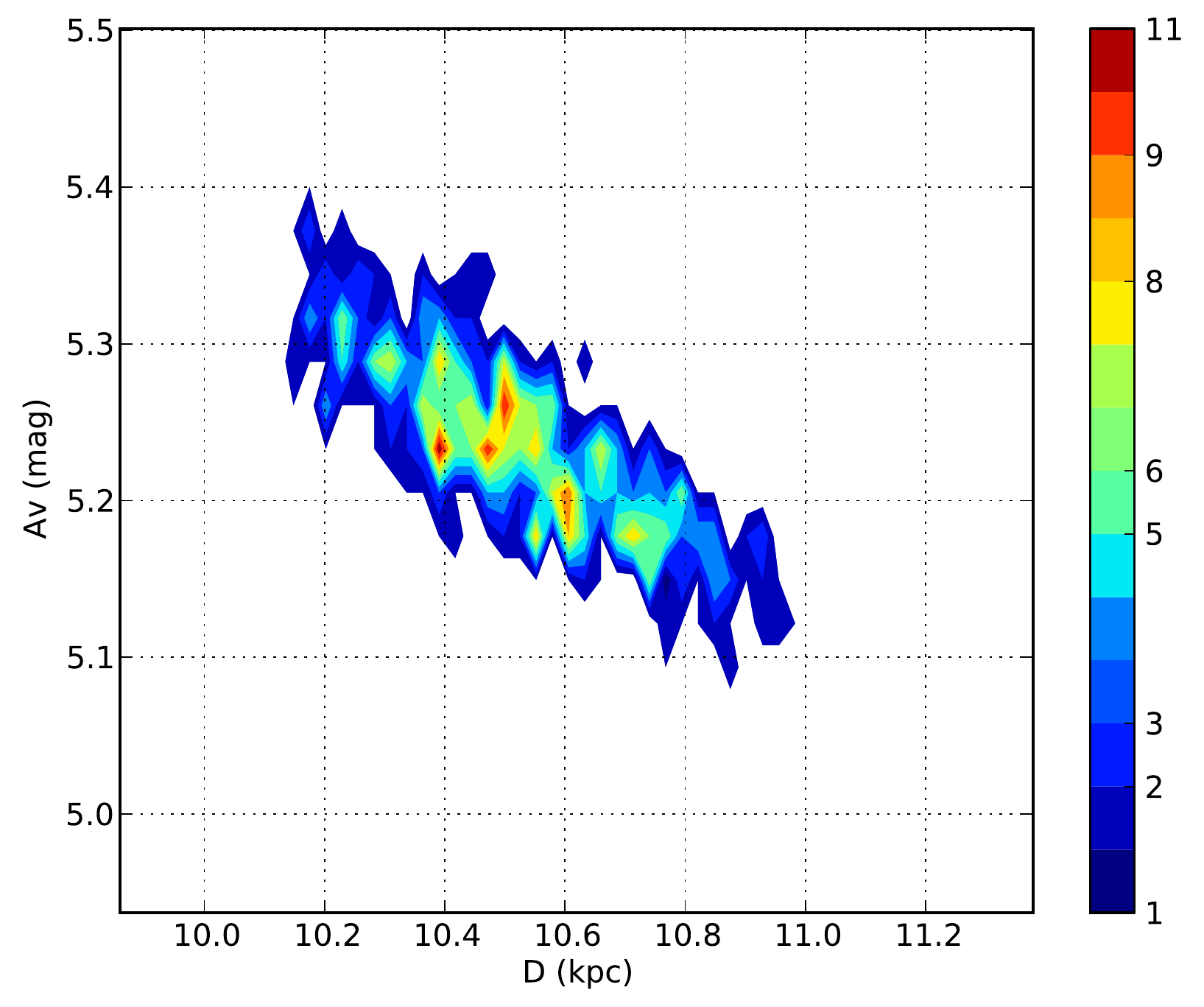}}
\caption{Result of Monte-Carlo simulations for the source 1E 1145.1-6141, to estimate uncertainties in D and A$_{\rm{V}}$ due to the fitting procedure. Colors represent the number of solutions as a function of  the two parameters (D and A$_{\rm{V}}$) values.}
\label{MCsimu}
\end{figure}

Degeneracy between several parameter values (based on the fitting procedure) needs to be taken into account. Indeed, solely relying on a single best fit does not capture the full phenomenology associated with SED fitting because D and $\rm{A}_{\rm{V}}$ are degenerate in this approach. In order to produce the best set of fits and to determine the dispersion on distance and extinction, we carried out 500 Monte Carlo simulations for each observed source, by varying the photometry within the uncertainties. Hence, we generate a random number from a normal distribution (assuming the photometric errors to be Gaussian), defined by the error bars for each photometric point, so that we build 500 new SEDs derived from the original one. These 500 new SEDs are subject to the same $\chi^2$ statistic computation as the one described above. Then, we have an entire set of best fits of parameters (D, $\rm{A}_{\rm{V}}$) and we are able to plot the distribution in the parameter space, showing the distribution of properties derived from these Monte-Carlo simulations, and especially showing the dispersion on distance and extinction for each source (see Figure \ref{MCsimu}). This dispersion value is taken as the error of the fitting procedure, and the median value of dispersion on distance determination for all considered HMXBs is then 0.75 kpc.\\

\begin{figure}[h]
\resizebox{\hsize}{!}{\includegraphics{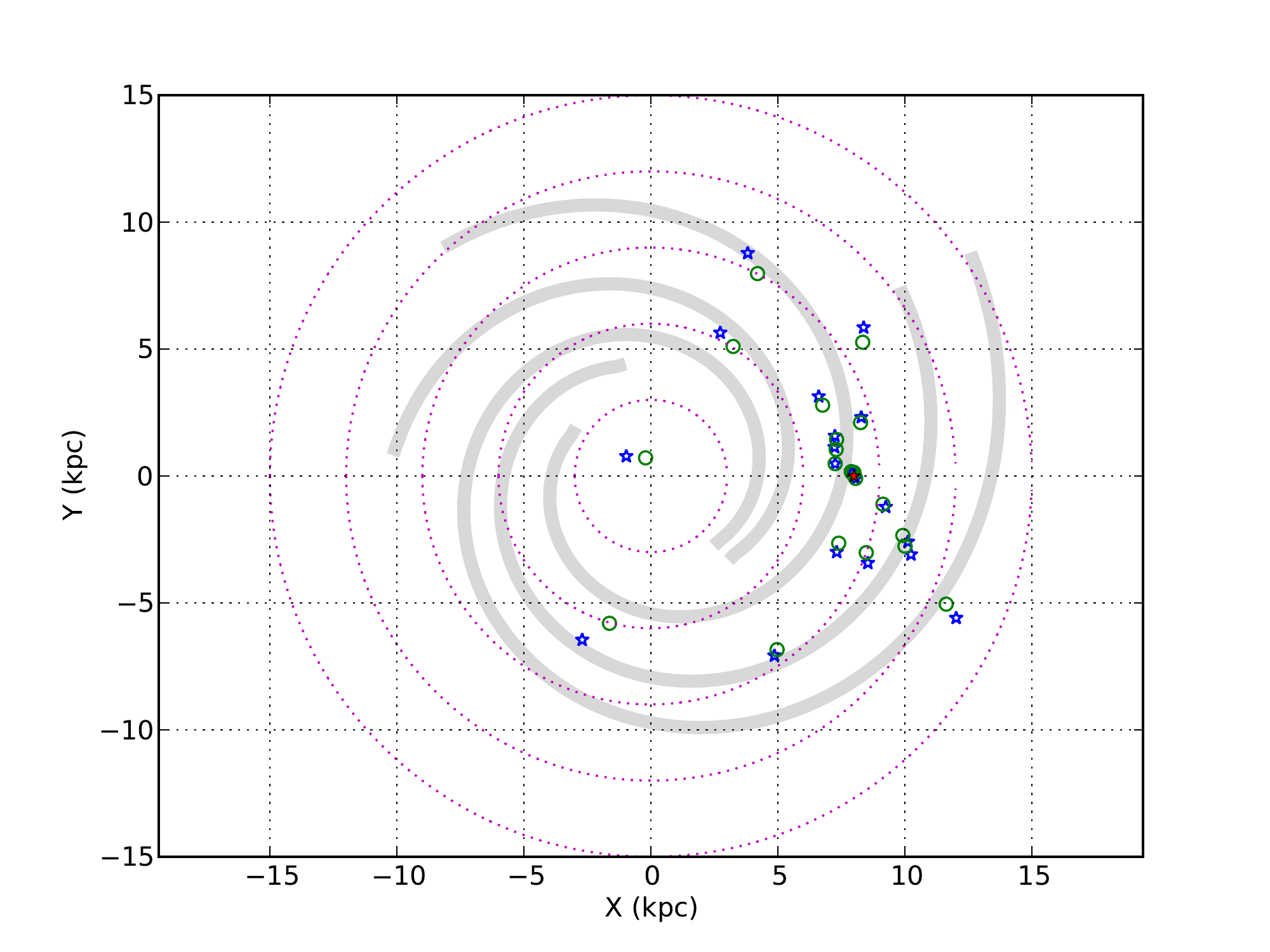}}
\caption{Green circles represent the initial positions of Be stars whereas blue stars represent the source positions taking into account the IR excess. Red star at (8.5 ; 0) represents the Sun location. Dotted pink circles represent radius of respectively 3, 6, 9, 12 and 15 kpc from the Galactic center.}
\label{excessIR}
\end{figure}

There are other sources of uncertainties, particularly the infrared excess of Be stars due to their circumstellar envelope generating free-free radiation. According to \citet{Dougherty_1994}, this excess should not exceed a mean 0.1 magnitude in J band, 0.15 magnitude in H band and 0.25 magnitude in K band. However, this value corresponds to absolute magnitude and therefore the excess can be smaller in apparent magnitude for sources located far away and higher for close ones. To take this effect into account, an estimate of the distance and extinction is needed. Since these two values are derived from the fitting procedure, we are only able to consider the distance and absorption values obtained without taking this IR excess into account. This approach is finally equivalent to adding a conservative error of 0.1, 0.15 and 0.25 magnitude to the apparent magnitude in the J, H and Ks bands respectively; this method presenting the advantage of treating all sources in an uniform way. Based on these results, presented in Figure \ref{excessIR}, we assume that this uncertainty will not affect the distances derived in this paper.\\

Furthermore, to test the potential additional error due to a different extinction law, we use $R_{\rm{V}}$ values given by \citet{Geminale_2004}. We affected to each HMXB an $R_{\rm{V}}$ value derived from the closest star present in their catalogue and we fitted the SED using the Cardelli law with the new $R_{\rm{V}}$ value. We find a median difference in distance determination of 0.07 kpc. Since deriving an accurate $R_{\rm{V}}$ value for each system appears to be observationaly biased (it depends on the closest star available in their catalogue), we decided not to take into account this additional uncertainty.\\

Finally, Table \ref{sample1} presents the 46 sources and their fundamental parameters (computed in this work or taken in the literature). Figure \ref{errors} represents most of the studied HMXBs with the uncertainties on their location computed taking into account all the errors described above: the error on distance due to uncertainties on the companion star radius and temperature and the error coming from the fitting procedure. We derived with the same way, the final error on $\rm{A}_{\rm{V}}$.

\section{Results: HMXB distribution and correlation with Star Forming Complexes}\label{correlation_method}

\begin{figure}[h]
\resizebox{\hsize}{!}{\includegraphics{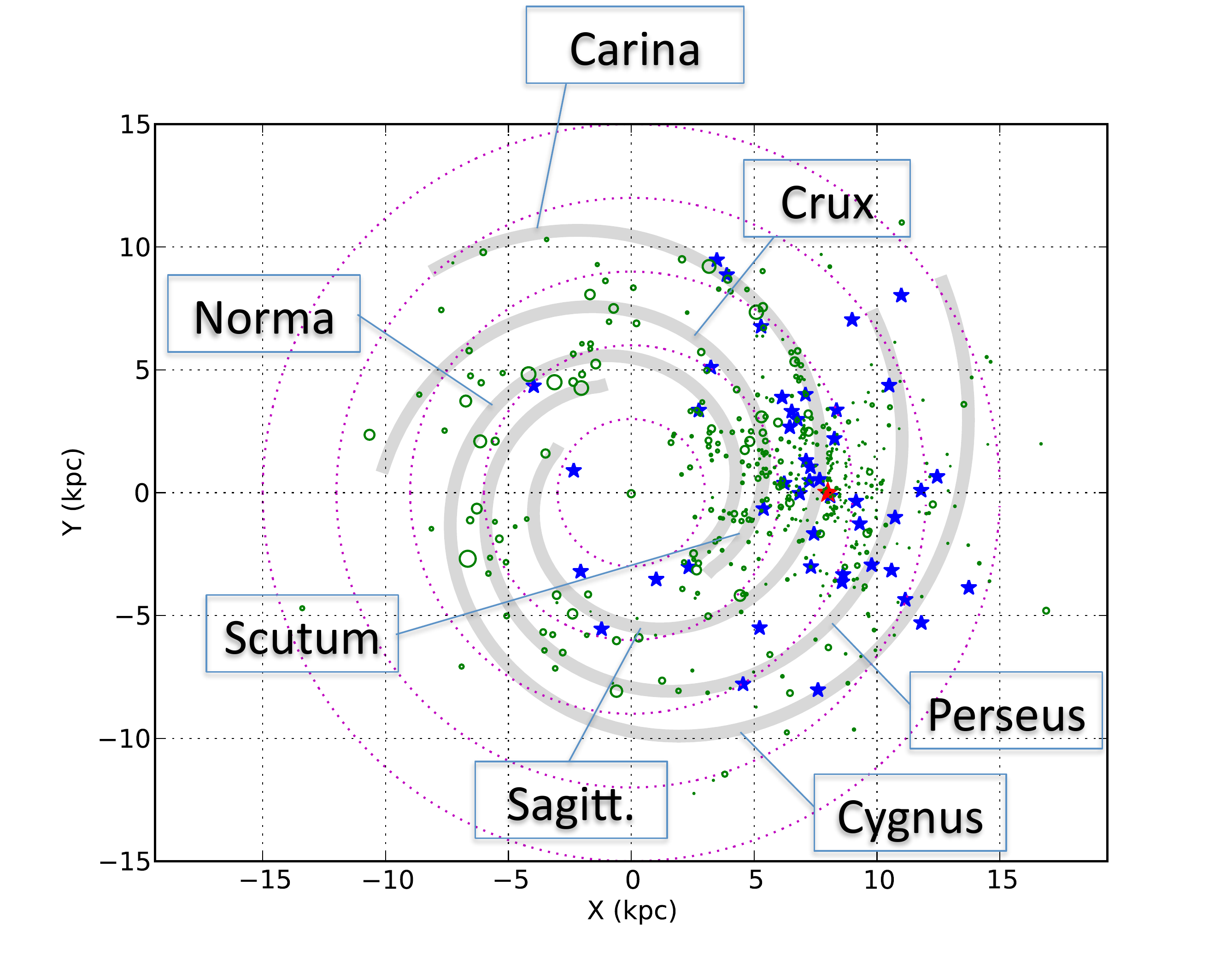}}
\caption{Distribution of HMXBs (blue stars) and SFCs (green circles). The circle radius of SFCs represents the excitation parameter value. The spiral arm model from \citet{Russeil_2003} is also plotted and the red star at (8.5 ; 0) represents the Sun position.}
\label{disti}
\end{figure}

We present in Figure \ref{disti} the distribution of HMXBs in the Galaxy, obtained with our novel approach based on distance determination. The spiral arms model given by \citet{Russeil_2003} is also presented. The question then arises: is there a correlation between this distribution of HMXBs and the distribution of Star Forming Complexes (SFCs) in the Milky Way (given by \citealt{Russeil_2003}), as it is expected from the short HMXB lifetime ? The first approach we adopt is to carry out a Kolmogorov-Smirnov test (KS-test) on each axis in order to quantify whether the two samples are drawn from the same probability distribution. We obtain a value of 0.15 for the X axis, a value of 0.25 for the Y axis and a value of 0.31 for the galactic longitude. These values are not negligible, suggesting that a correlation between the two samples does exist, though part of the information is lost because of the projection on the two axis. To overcome this caveat, we propose another method described hereafter.\\

\begin{figure}[h]
\resizebox{\hsize}{!}{\includegraphics{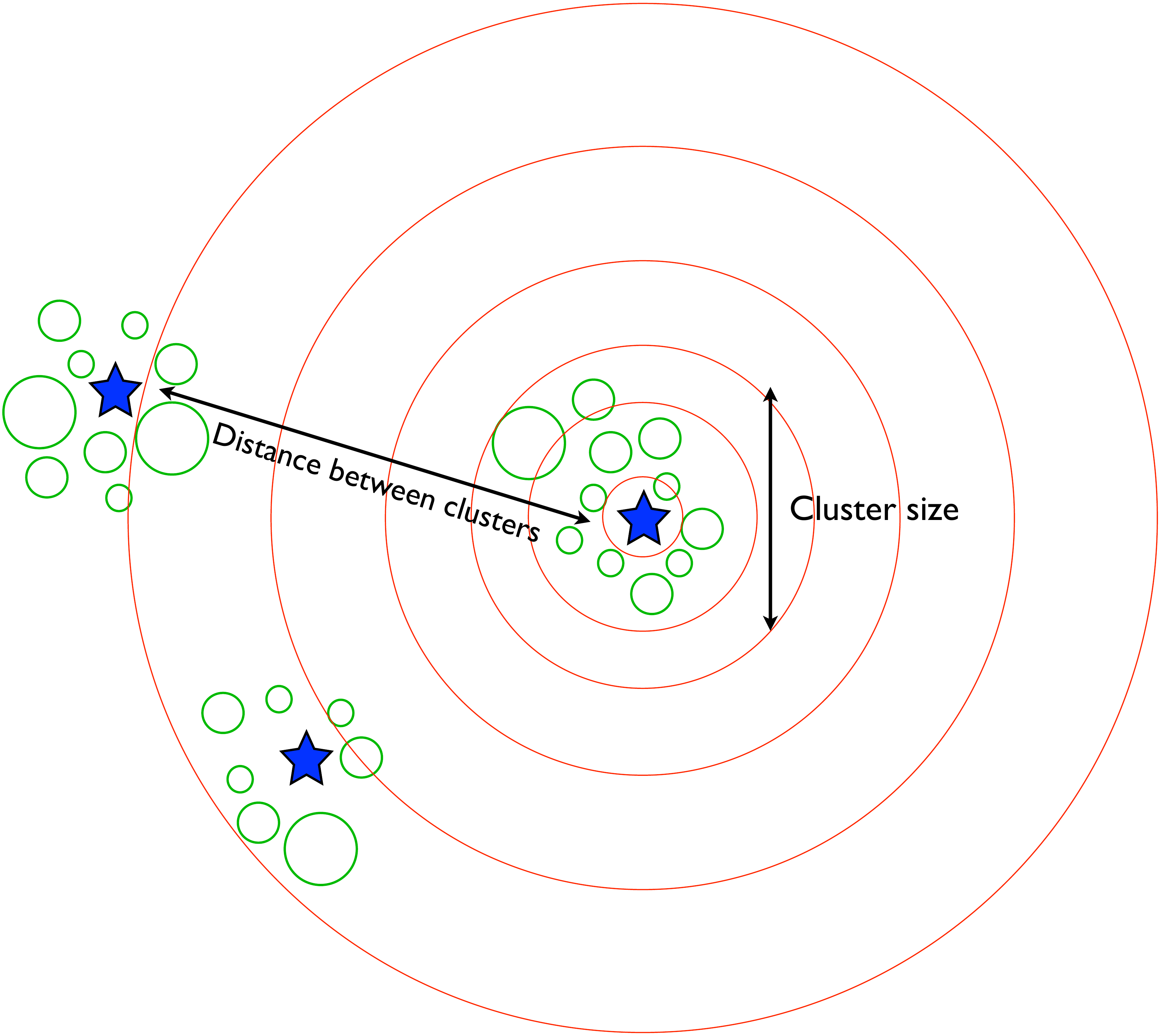}}
\caption{Description of the method used to evaluate the correlation between HMXBs and SFCs.}
\label{methode}
\end{figure}

We assume that each HMXB (blue stars on Figure \ref{methode}) is clustered with several SFCs (green circles). Hence, we can define two characteristic scales: a typical cluster size and a typical distance between clusters. Around each HMXB, we define several circles with different radii (red circles) and we finally count the number of HMXBs for which at least one SFC is within the specified radius (called `number of correlations" hereafter).

\begin{figure}[h]
\resizebox{\hsize}{!}{\includegraphics{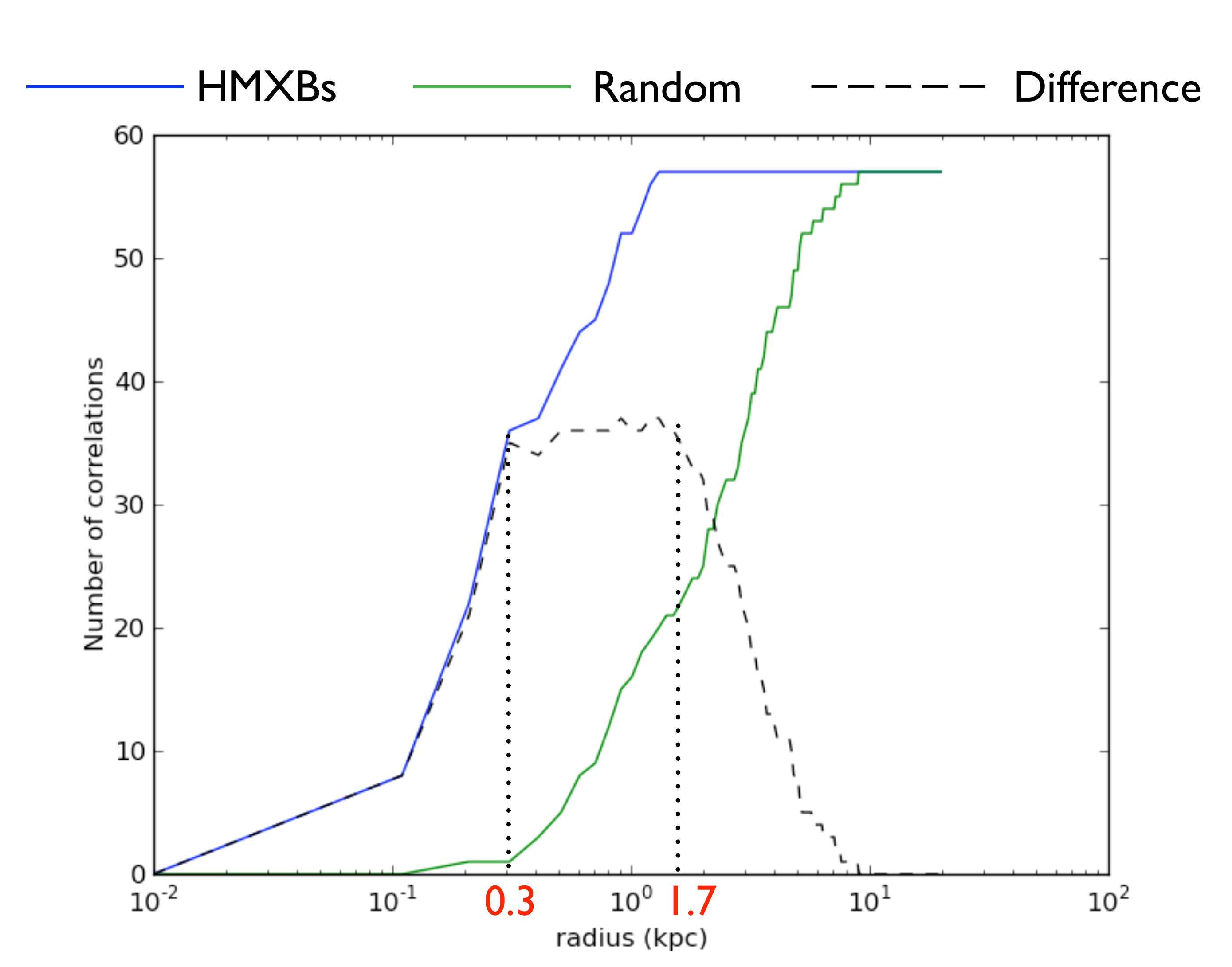}}
\caption{Result of the correlation determination in 2D.}
\label{correl_result}
\end{figure}

\begin{figure}[h]
\resizebox{\hsize}{!}{\includegraphics{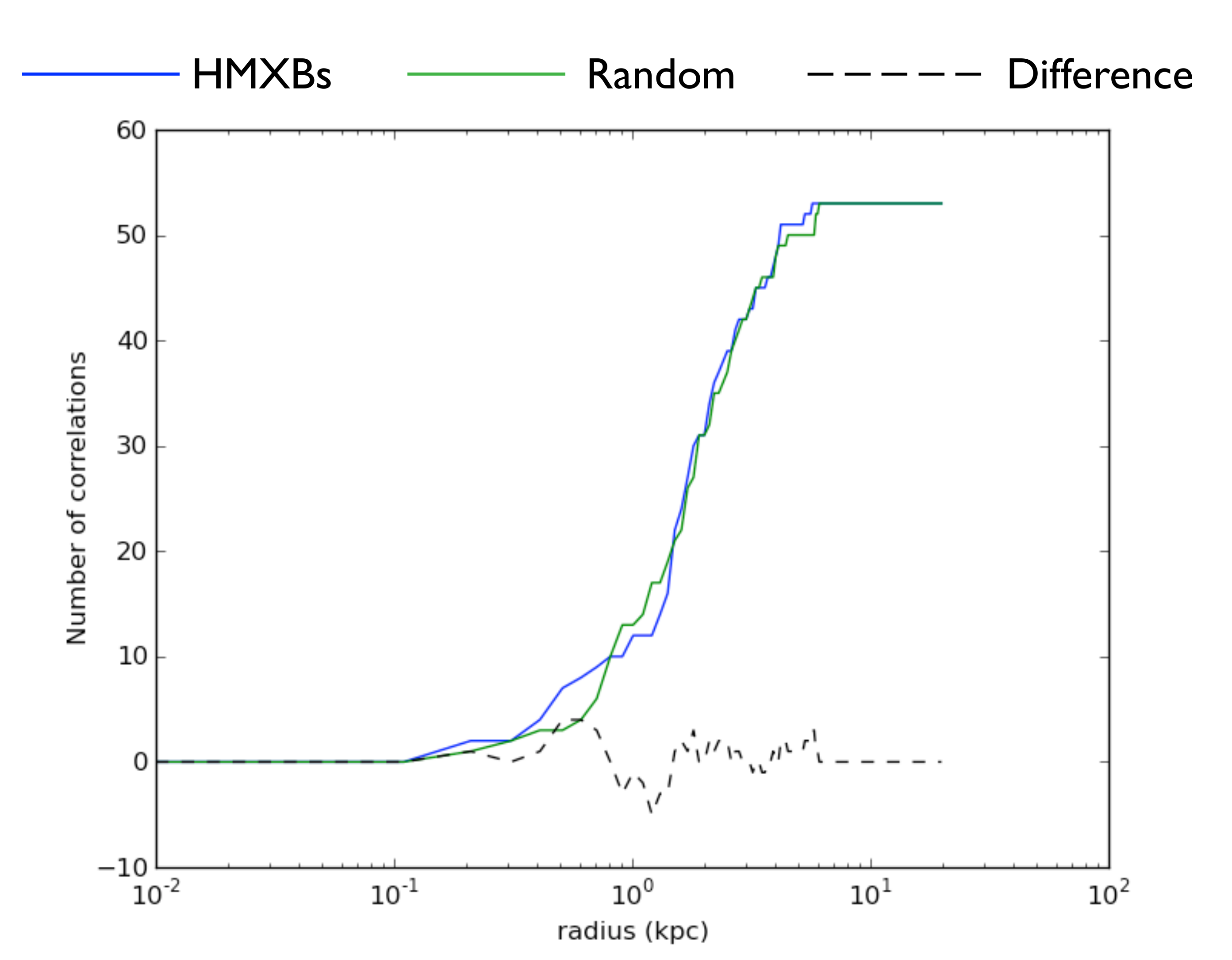}}
\caption{Result of the correlation determination with globular clusters.}
\label{GC_correl}
\end{figure}

The number of correlations versus the circle radius is plotted as the blue curve on Figure \ref{correl_result}. The green curve is that expected from chance correlations assuming the HMXBs are evenly distributed across the sky following an uniform distribution on x and y axis between 0 and 15 kpc. The dashed line represents the difference between the two previous curves. This difference, being non equal to zero, allows us to state that a strong correlation exists between HMXB and SFC positions in the Milky Way. Moreover, we compute the two characteristic scales described above: the typical cluster size of 0.3 kpc and the typical distance between clusters of 1.7 kpc with uncertainties of 0.05 kpc and 0.3 kpc respectively, using the uncertainty on HMXB distance of 17\% as derived in Section \ref{fit}.  \\
If we take into account the uncertainties in HMXB positions (described in Section \ref{unc_section}) and in SFC positions (given in \citealt{Russeil_2003}, median error of 0.25 kpc), the correlation still exists with the same cluster size and the same distance between clusters. \citet{Bodaghee_2012} mention that HMXBs and OB complexes are clustered for a cluster size $r < 1$ kpc. 
This upper limit, obtained with a different method, is consistent with our results. Finally, we test our correlation code using a sample of globular clusters (\citealt{Bica_2006}), principally located in the Galactic bulge. Figure \ref{GC_correl} shows the result of correlation test. The number of correlation as a function of the circle radius (blue curve) follows the trend of the green curve, showing the evolution expected from chance correlation, when assuming that HMXBs are evenly distributed across the sky. Clearly, as expected, no correlation is seen, assessing the robustness of our correlation evaluation method.

\section{Implication of the correlation on HMXB formation and evolution}

The distribution of HMXBs reflects the stellar formation that took place some tens of Myr ago, since they are not an instantaneous star formation rate (SFR) indicator as explained in \citet{Shtykovskiy_2007}. Then, an offset between spiral arms (an indicator of the actual star formation) and HMXB distribution is expected. Indeed, since the spiral arm rotation velocity is different than the angular velocity of the stellar disk, we expect HMXB positions to be offset from the currently visible SFCs. This time lag was mentioned in \citet{Lutovinov_2005} and \citet{Dean_2005} but a deeper investigation of this issue was not possible due to the limited sample of HMXBs at that time. \citet{Shtykovskiy_2007} evaluated the offset for the galaxy M51. Here, we attempt to implement this formalism in the case of the Milky Way with the approach we have mentioned in \citet{Coleiro_2011} .\\

One can suppose the spiral arm density wave to rotate with the speed $\Omega_p$ = 24 km\,s$^{-1}$\,kpc$^{-1}$ (see e.g. \citealt{Dias_2005}), in the same direction as the stellar disk, which velocity curve is assumed to be flat in the galactocentric distance range of interest, according to \citet{Brand_1993}: 

\begin{equation}
V/(220 \rm{km.s}^{-1})=a_1(\mathit{r}/8.5 kpc)^{a_2}+a_3 ,
\label{Brand}
\end{equation}

with $\rm{a}_1$=1.00767, $\rm{a}_2$=0.0394, $\rm{a}_3$=0.00712 and $r$ is the radius from the Galactic center.

\begin{figure}[h!]
\resizebox{\hsize}{!}{\includegraphics{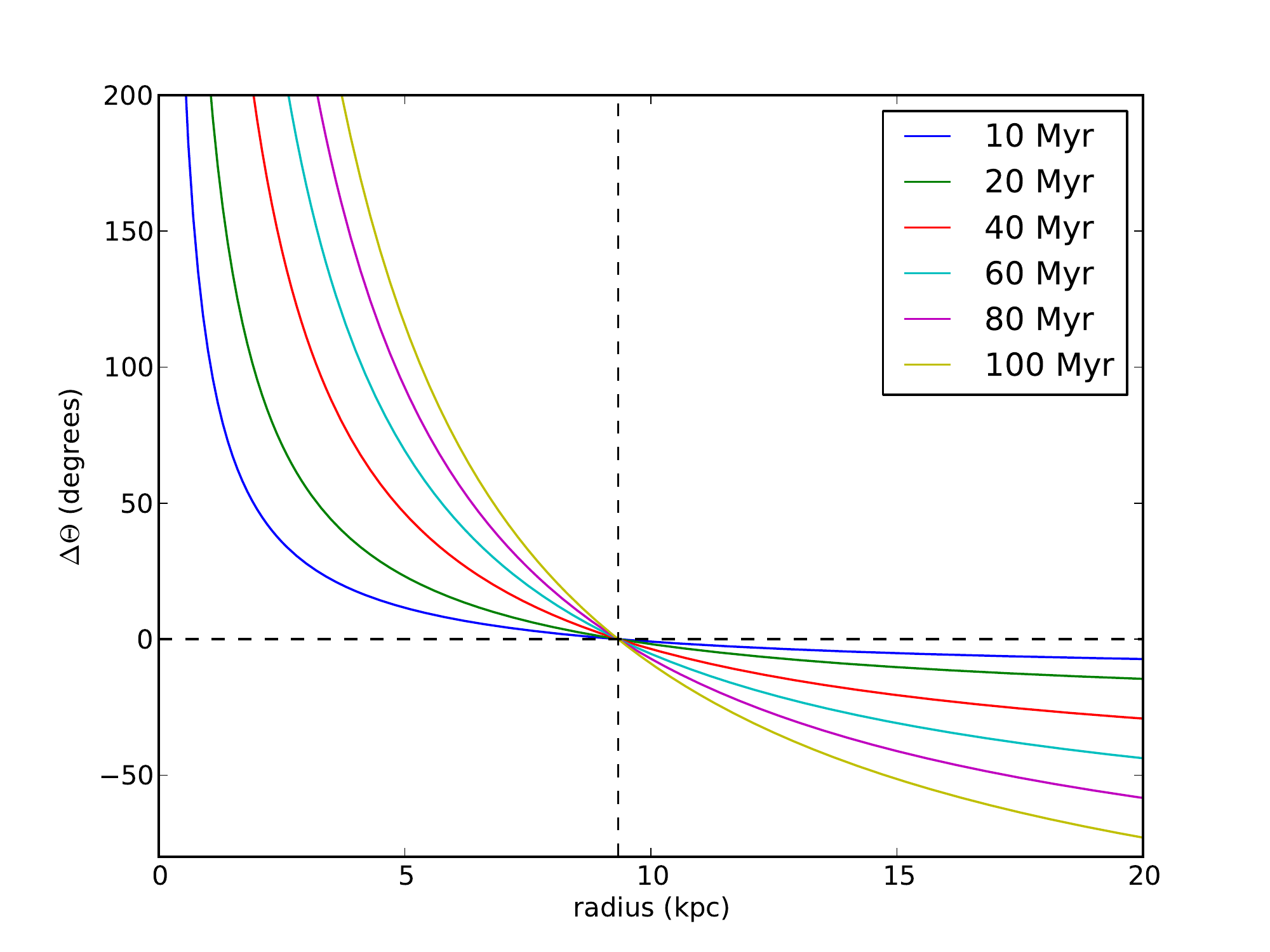}}
\caption{Evolution of the offset angle between spiral arms and expected positions of sources as a function of the radius in the Galaxy according to the age of sources. The vertical dashed line indicates the corotation radius.}
\label{angle}
\end{figure}

\begin{figure}[h!]
\resizebox{\hsize}{!}{\includegraphics{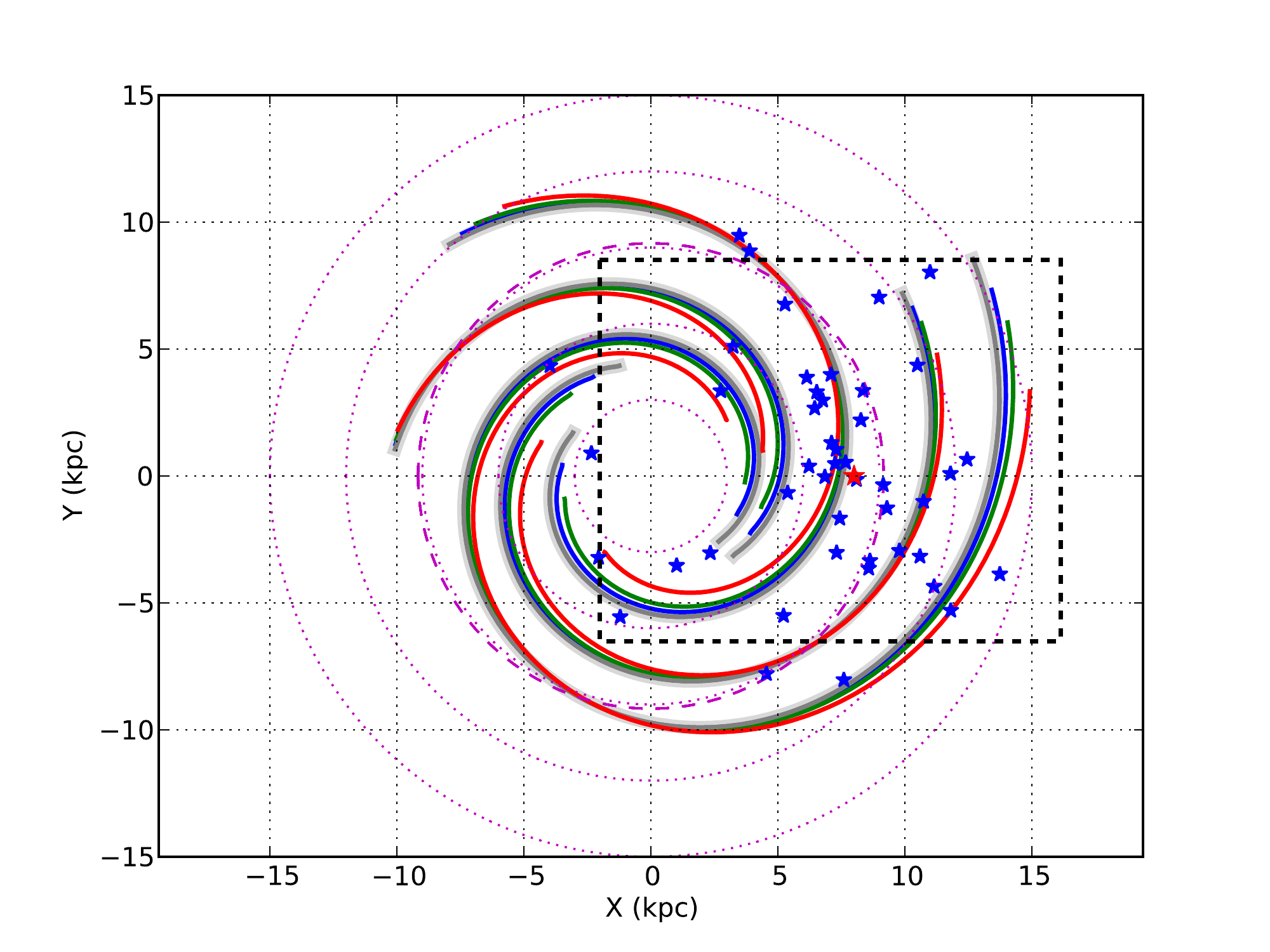}}
\caption{Distribution of HMXBs and spiral structure. The blue, green and red curves correspond respectively to the expected locations of objects with ages of 10, 20 and 40 Myr computed with Equation \ref{dT}. For better clarity, we do not represent expected locations of objects with ages of 60, 80 and 100 Myr. The grey curve represents the current position of spiral arms and the dashed circle indicates the corotation radius. Dotted circles represent radii of 3, 6, 9, 12 and 15 kpc. Dashed frame represents the region shown in Figure \ref{errors}.}
\label{spiral_struct}
\end{figure}

Then, to a first approximation, it is possible to locate the expected HMXB locations relative to the current position of the spiral arms in time $\tau$, i.e., the angular offset $\Delta\Theta(r)$ by the equation:

 \begin{equation}
\Delta\Theta(r)=\left(\Omega(r)-\Omega_p\right)\tau ,
\label{dT}
\end{equation}

where $\Omega(r)$ is the galactic rotation curve derived from Equation \ref{Brand}.
To estimate the displacement of HMXBs relative to the current position of the spiral density wave, we plot the angular offset, $\Delta\Theta(r)$, as a function of the radius for sources with different ages (10, 20, 40, 60, 80 and 100 Myr.) (Figure \ref{angle}). In a second time, we plot the expected positions of sources formed 10, 20, 40 Myr ago in the Galaxy map (Figure \ref{spiral_struct}) (For better visibility, here, we do not plot the expected positions of sources formed 60, 80 and 100 Myr ago).\\

Even if Figures \ref{spiral_struct} and \ref{errors} do not visually suggest any substantial offset between the current spiral arm position and the expected position of HMXBs (which depends on the age of the sources) we would now like to quantitatively assess the presence of an offset.

\subsection{Existence of an offset between HMXBs and Galactic spiral arms}\label{dist_calc}

To perform this study we need to consider a number of different issues. As underlined above, the expected offset depends on the age of the X-ray sources, hence, to highlight this offset, we must split the sample of HMXBs depending on the age of the sources. Two different samples are then created: one containing 4 supergiant stars (luminosity classes I or II according to \citealt{Charles_2006}) and a second one containing 9 Be stars (luminosity class III or V according to \citealt{Charles_2006}). We explain the way these samples were created in Section \ref{ageHMXBs} . 

\begin{figure*}[h!]
\resizebox{\hsize}{!}{\includegraphics{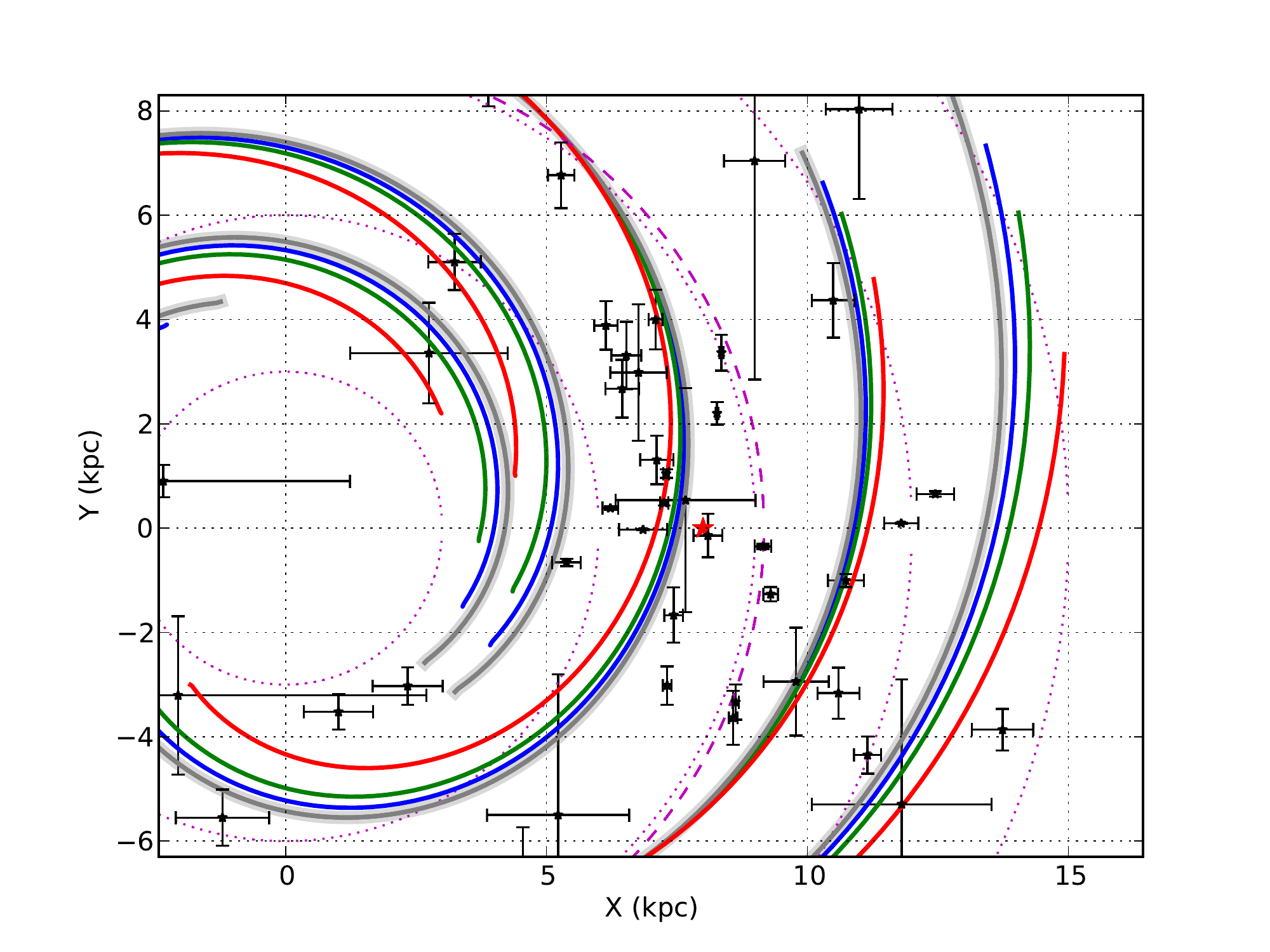}}
\caption{Positions of HMXBs with error bars (zoom in the Solar region defined by dashed frame in Figure \ref{spiral_struct}).}
\label{errors}
\end{figure*}

We calculate the distance from each HMXB to the closest actual spiral arm (given in \citealt{Russeil_2003}) using:

\begin{equation}
\rm{distance}=\sqrt{(X_{\rm{source}}-X_{\rm{arm}})^2+(Y_{\rm{source}}-Y_{\rm{arm}})^2} ,
\label{distance_form}
\end{equation}

where ($X_{\rm{source}}$, $Y_{\rm{source}}$) are the coordinates of the source and ($X_{\rm{arm}}$, $Y_{\rm{arm}}$) are the coordinates of the closest point on the arm. 

\begin{figure*}[h!]
\resizebox{\hsize}{!}{\includegraphics{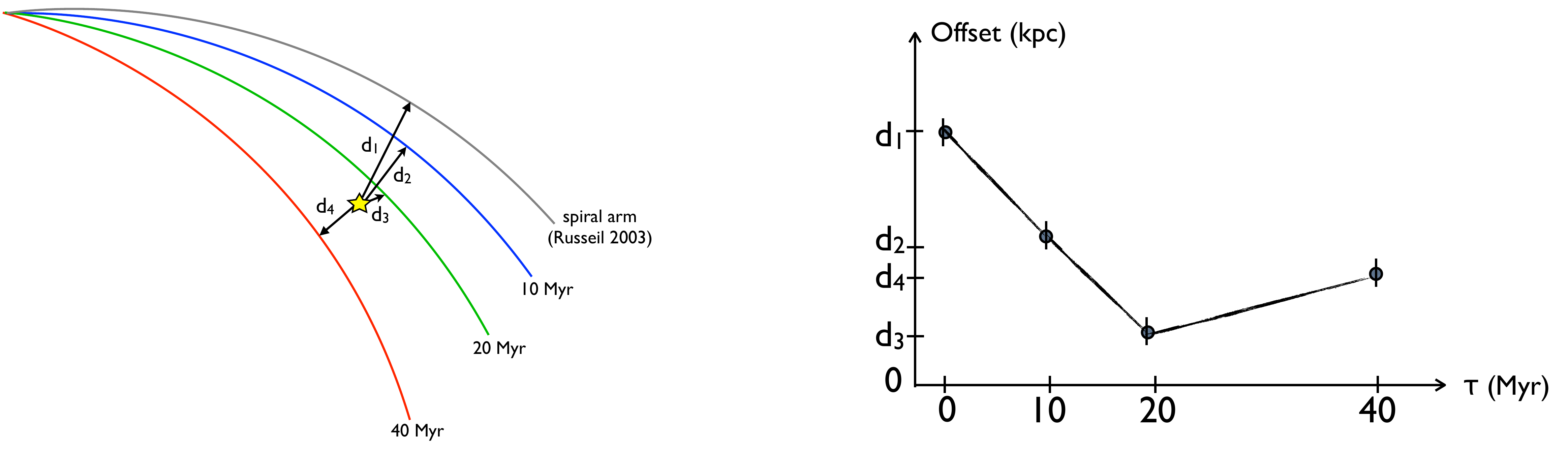}}
\caption{Method used to derive the age of a sample of HMXBs. As in Figure \ref{errors}, a spiral arm is represented with the expected positions of sources of 10, 20 and 40 Myr old. An HMXB is represented by the star and distances to current spiral arm and to each expected positions are called d$_1$, d$_2$, d$_3$ and d$_4$ respectively. The evolution of this distance as a fonction of the time is represented in the diagram on the right. In this case, the source would be $\sim$ 20 Myr old.}
\label{distances_expl}
\end{figure*}

\begin{figure}[h!]
\resizebox{\hsize}{!}{\includegraphics{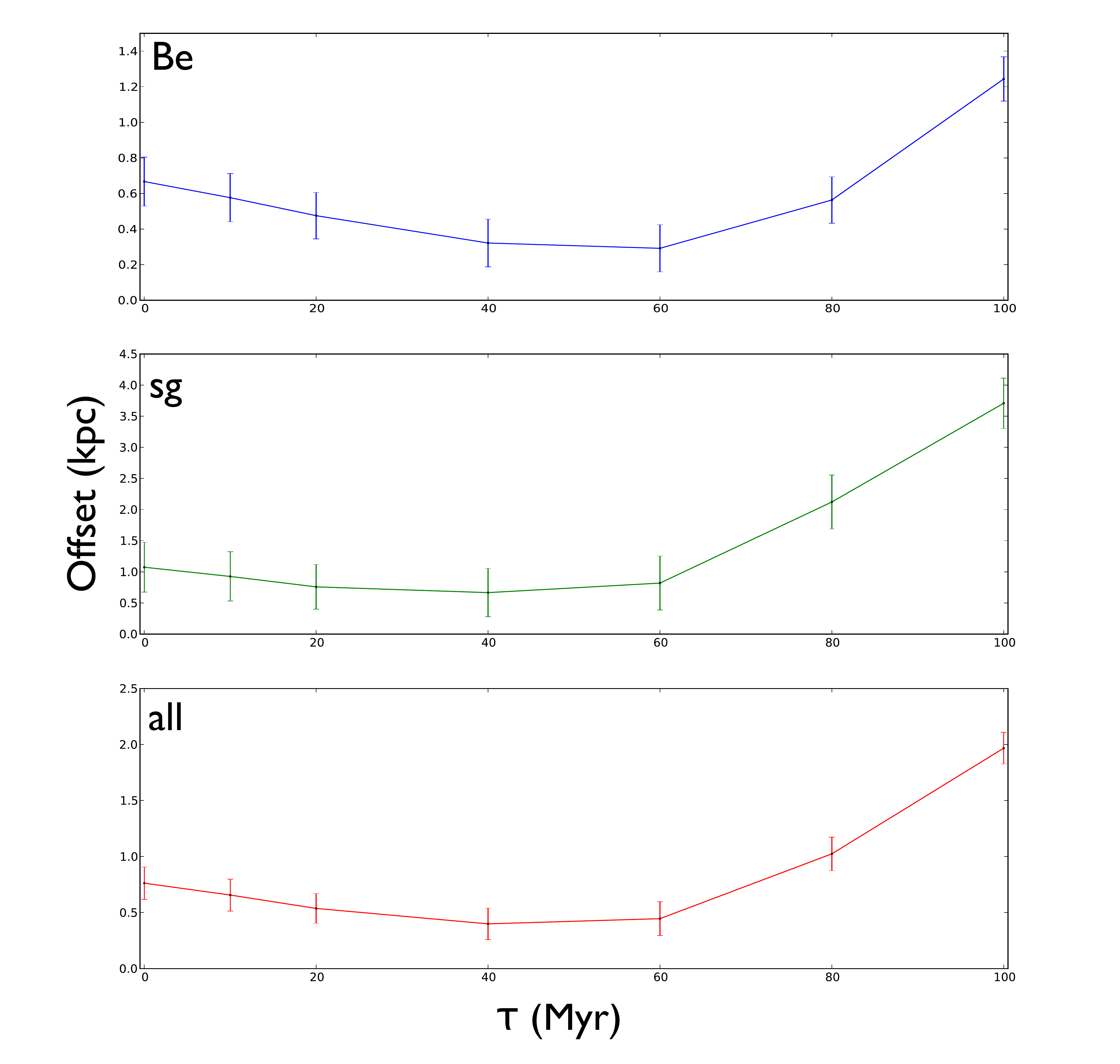}}
\caption{Evolution of the mean distance (in kpc) between source and closest expected position versus time $\tau$ (in Myr) for Be sources (upper panel), supergiant sources (mid panel) and all the sources (lower panel). Error bars are 1-$\sigma$ uncertainties.}
\label{evol_offset}
\end{figure}

We follow exactly the same procedure to calculate the distance from each source to the closest expected position of sources formed 20, 40, 60, 80 and 100 Myr ago and we determine the mean value of the offsets (taking into account all the sources of the two samples). The sources are expected to be closest to one of the expected positions computed above than to the current spiral arms observed by \citet{Russeil_2003}. The method is described in Figure \ref{distances_expl} (for instance, in Figures \ref{errors} and \ref{distances_expl}, a source of 20 Myr should be located closer to the green arm representing the expected position of a 20 Myr-old HMXB than to the current spiral wave position). Results are given in Figure \ref{evol_offset}.

We determine the error bars using the propagation of uncertainty formula. The 1-$\sigma$ error associated to the distance between each source $i$ and the closest point on the arm is then given by the following equation:

\begin{equation}
\begin{split} 
\sigma_{i}^2 = & \frac{(X_{\rm{source}}-X_{\rm{arm}})^2\left[\sigma_{X_{\rm{source}}}^2+\sigma_{X_{\rm{arm}}}^2\right]}{(X_{\rm{source}}-X_{\rm{arm}})^2+(Y_{\rm{source}}-Y_{\rm{arm}})^2} \\ 
 & +\frac{(Y_{\rm{source}}-Y_{\rm{arm}})^2\left[\sigma_{Y_{\rm{source}}}^2+\sigma_{Y_{\rm{arm}}}^2\right]}{(X_{\rm{source}}-X_{\rm{arm}})^2+(Y_{\rm{source}}-Y_{\rm{arm}})^2}
\end{split} 
\label{error_arm_HMXB}
\end{equation}

where ($\sigma_{X_{\rm{source}}}$ , $\sigma_{Y_{\rm{source}}}$) are the errors on source position detailed in Section \ref{unc_section} and ($\sigma_{X_{\rm{arm}}}$ , $\sigma_{Y_{\rm{arm}}}$) are the errors on arm position taken to be equal to zero. Finally, the uncertainty associated to the mean value is calculated as follows:

\begin{equation}
\sigma=\frac{1}{N}\sqrt{\sum_{i}^{N}\sigma_{i}^2}
\end{equation}

where $N$ is the number of sources and $\sigma_{\rm{i}}$ is the error associated with the i$^{\rm{th}}$ source.\\

Taking into account all the sources of the sample (Figure \ref{evol_offset}, lower panel), we cannot observe any significant variation of the offset with time before 60 Myr. This result was expected because the minimum peak of distance between each HMXB and the closest expected position should be clearly different whether we consider Be or supergiant stars. Also, by taking all spectral types into account, we tend to loose a part of the information except the mean age upper limit of 60 Myr highlighted by the plot. For both supergiant and Be samples, we observe a more significant increase of the offset after 60 Myr (Figure \ref{evol_offset}, first and mid panels). This increase could be a signature of the expected offset between the HMXB positions and the current spiral arms. However, we must be cautious about this result especially because of the small number of HMXBs in the two samples.

Even if the locations of the sources are accurately determined, several reasons may affect the HMXB density and prevent the offset detection as underlined by \citet{Lutovinov_2005}: complex motion of density wave and stars from their birthdate to the X-ray phase, presence of previously undetected parts of the Galactic spiral arms, observational selection effect, etc. Finally, a larger sample of supergiant type HMXBs is needed to confirm the offset detection more confidently.\\

Moreover, the time interval during which HMXBs appear should translate the mass range of both stars of binary systems (see \citealt{Dean_2005}). The results presented on Figure \ref{evol_offset} only enable us to state that this time interval is lower than 60 Myr on average for all stars. 

\subsection{Deriving the age of HMXBs}\label{ageHMXBs}

Following the method described above, one can compute a distance between each source and the different theoretical arms that correspond to the current predicted position of a source sample born 20, 40, 60, 80 or 100 Myr ago. In the previous section we studied a sample of Be and supergiant HMXBs, and this could also be applied to each source separately, to constrain the age and the potential migration of the system due to a supernova kick.\\

We choose not to take into account here the arm width that could be translated as an uncertainty on the position of the expected position of sources. Indeed, by taking this position dispersion into consideration, 1-$\sigma$ errors considerably increase and prevent any conclusion. Then, we assume all sources to be formed at the central position of the arm.\\ 

We expect the distance from the source to the expected position to decrease until the expected position corresponds to the age of the source, and then to increase afterwards (see Figure \ref{distances_expl}). For some sources, the offset change is more complicated and can be explained as follows: 

\begin{enumerate}
\item belonging to one of the four arms of the Milky Way is not well established when the distance between a source and the theoretical expected positions is seen as always increasing with time;

\item for sources located close to the corotation radius, current spiral arms and expected positions of HMXBs of different ages are almost superimposed explaining the quasi-constant distance to theoretical positions observed for some sources. Since offsets between expected positions of sources of different ages are very small, it can be possible to ascribe an incorrect age to a source located close to an expected position that does not correspond to its real age;

\item sources that underwent more significant kicks could be incorrectly associated with a theoretical position of sources of different age.

\end{enumerate}

Then, for this study, we only take into account sources presenting an expected offset evolution during time, to make the conclusions easier. 

\begin{figure*}[h!]
\resizebox{\hsize}{!}{\includegraphics{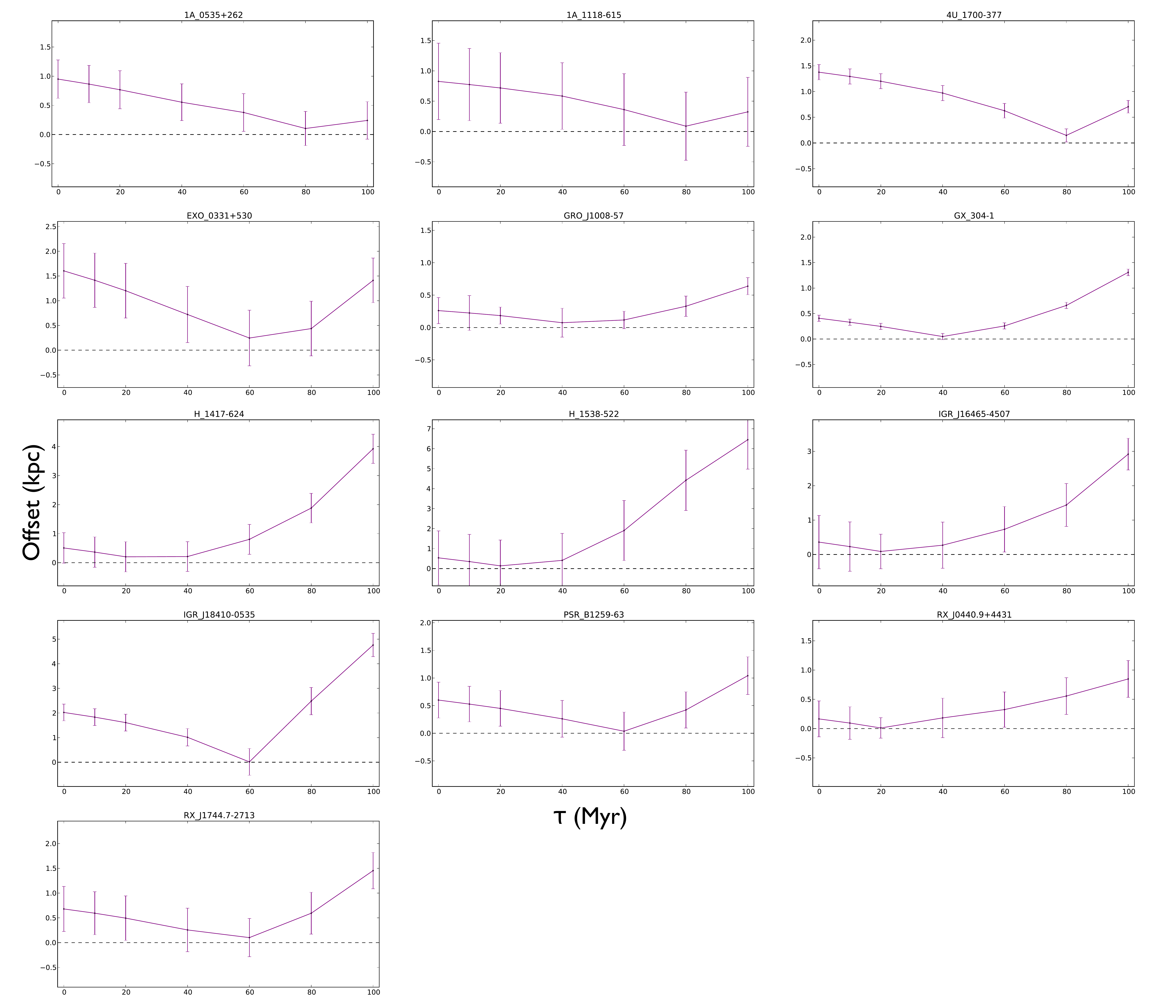}}
\caption{Evolution of the distance between observed and expected positions (in kpc) versus time $\tau$ (in Myr) for individual HMXBs. Error bars are 1-$\sigma$ uncertainties.}
\label{evol_offset_all}
\end{figure*}

We present on Figure \ref{evol_offset_all} results for BeHMXBs and sgHMXBs for which the evolution of the offset as a function of time is consistent with the theory. It is possible to determine the rough age of these sources and a lower and upper limit of kick migration distances. The results are summarized in Table \ref{agemigration} and will be commented in the following paragraphs.\\

\begin{center}
\begin{table}
\scalebox{0.77}
{
\begin{tabular}{lrrr}

SOURCE NAME & AGE (Myr) & MIGRATION DISTANCE (kpc) & UNCERTAINTY\\
\hline
\multicolumn{4}{c}{\textbf{Be}}\\
1A 0535+262 & 80 &	       0.10	 &     0.30      \\
1A 1118-615 & 80 &	       0.088  &    0.56  \\   
EXO 0331+530 & 60 &	       0.25	  &    0.080\\     
GRO J1008-57 & 40 &	        0.074	  &    0.15\\     
GX 304-1 & 40 & 	       0.048	  &    0.59 \\    
H 1417-624 & 20 & 	       0.20	  &    0.39  \\   
PSR B1259-63 & 60 & 	       0.037  &    0.51  \\   
RX J0440.9+4431 & 20 & 	       0.011  &    0.17  \\   
RX J1744.7-2713 & 60 & 	       0.10  &    1.0  \\   

\hline
\multicolumn{4}{c}{\textbf{Supergiants}}\\
4U 1700-377 & 80 &	 0.15	  &    0.28\\     
IGR J16465-4507 & 20 & 	       0.087	  &    0.052    \\
IGR J18410-0535 & 60 & 	       0.013  &    0.11  \\   
H 1538-522 & 20 & 	       0.14	  &    0.52    \\ 
\hline

\end{tabular}
}
\caption{Age and migration distance derived respectively for BeHMXBs (top) and sgHMXBs (bottom). Uncertainties are derived by Equation \ref{error_arm_HMXB}, i.e. correspond in Figure \ref{distances_expl} to the error on the minimum distance d$_{3}$.}
\label{agemigration} 
\end{table}
\end{center}

If we consider only the sources studied separately here (sources showing the expected evolution of offset as a function of age), we can derive a mean age of $\sim$ 45 Myr for sgHMXBs and 51 Myr for BeHMXBs. This result seems consistent with the distinct evolution timescales of these two kinds of systems but again, we must be cautious about this result because of the small sample of sources used in this calculation (4 supergiant and 9 Be systems).

\subsection{Constraints on migration from supernova kicks}

Even if the theory is still poorly understood, it is admitted that a small asymmetry in the way a supernova explodes could make a binary system move due to a substantial increase of its velocity in a particular direction. We computed the cluster size (in the same way as in Section \ref{correlation_method}) only for Be stars that are expected to have experienced a supernova kick event during their evolution and we found a value of 0.3 kpc, consistent with the migration distances underlined in \citet{Bodaghee_2012}. 
This value also enables us to constrain the time lapse between supernova event and HMXB stage, assuming a regular value of kick velocity of 100 km\,s$^{-1}$ (for higher kick velocity, fewer system should remain bound, \citealt{Hills_1983}) and a maximum migration distance due to the kick event of 0.3 kpc, we reach an upper limit for the time lapse between the supernova explosion and the HMXB step of around 3 Myr.\\

To improve this study we focus on the sources selected before (4 sgHMXBs  and 9 BeHMXBs, Table \ref{agemigration}). The distance between the object and the closest expected position gives a kick value for each source and a mean value depending on the spectral type of the two samples. Results are presented in Table \ref{agemigration}. We derive a mean migration distance of 0.11 kpc for BeHMXBs and of 0.10 kpc for sgHMXBs. Again, we should be careful about these mean values because of large error bars and small samples. Moreover, it is important to underline that these derived values only represent a lower limit to the kick migration distance since we cannot take into account the migration distance on galactic latitude given by the kick. Then, we only get the projected migration distance on the Galactic plane.

\section{Conclusion}\label{conclusion}
Examining the distribution of HMXBs is of major interest in order to study in depth the formation of these high energy sources. However, HMXB locations are usually poorly constrained and largely dependent on the determination method. Here, for the first time, we determine the location of a sample of HMXBs using an uniform and accurate approach: SED fitting of their distance and absorption. This method, based on a least-square minimization, enables us to reveal a consistent picture of the HMXB distribution, following the spiral arm structure of the Galaxy. The uncertainties lead to a small error on source location and allow us to tackle the study of the correlation with SFC distribution. This study shows that HMXBs are clustered with SFCs and enables to quantitatively define the cluster size (0.3 $\pm$ 0.05 kpc) and the distance between clusters (1.7 $\pm$ 0.3 kpc). We go further by quantitatively assessing the offset between current spiral density wave position and expected HMXB positions due to the fact that the matter rotation velocity is different than the spiral arm rotation speed. Exploring the environment in which such binary systems were formed is of major interest to study the properties of these binary systems such as stellar mass, dust cocoon density, etc. 
Here, we quantitatively show the correlation between HMXB distribution and Star Forming Complexes distribution. Even if we highlighted the expected offset between current spiral arms and source positions for some sources, it remains difficult to assess it for the entire sample. Undoubtedly, this assessment will be improved using a larger sample of sources, accurate Galactic spiral arms model and a dynamical model of matter and density wave. Our method of investigation does not give exhaustive results for the entire sample of sources because several sources are located close to the corotation radius and, for some sources, an association with one of the four arms of the Milky Way is not well established. However, for 4 sgHMXBs and 9 BeHMXBs, we are able to derive an age (mean age of 51 Myr for BeHMXBs and 45 Myr for sgHMXBs) and a migration distance (mean value of 0.11 kpc for BeHMXBs and 0.10 kpc for sgHMXBs), giving constraints on the supernova explosion kick. This study represents an important progress in the investigation of formation and evolution of these binary systems.

\begin{acknowledgements}
We warmly thank the anonymous referee for constructive comments which allowed us to improve the manuscript. We are pleased to thank P. A. Curran for his careful rereading of the paper. We acknowledge A. Bodaghee, P.A. Charles, P.A. Curran, C. Knigge, F. Rahoui, M. Servillat and J.A. Zurita Heras for useful discussions. This work was supported by the Centre National d'Etudes Spatiales (CNES), based on observations obtained with MINE --the Multi-wavelength INTEGRAL NEtwork--. This research has made use of the IGR Sources page maintained by J. Rodriguez \& A. Bodaghee (http://irfu.cea.fr/Sap/IGR-Sources/); of data products from the Two Micron All Sky Survey, which is a joint project of the University of Massachusetts and the Infrared Processing and Analysis Center/California Institute of Technology, funded by the National Aeronautics and Space Administration and the National Science Foundation; of the SIMBAD database and the VizieR catalogue access tool, operated at CDS, Strasbourg, France as well as of NASA's Astrophysics Data System Bibliographic Services.
\end{acknowledgements}

  \bibliographystyle{apj} 
\bibliography{biblio_1}

\end{document}